\documentclass[reprint,amsmath,amssymb,aps]{revtex4-2}
\usepackage[colorlinks,citecolor=blue,linkcolor=blue,anchorcolor=blue,filecolor=blue,
urlcolor=blue]{hyperref}
\usepackage{graphicx}
\usepackage{ulem}
\usepackage{dcolumn}
\usepackage{xurl}
\usepackage{bm}
\usepackage{todonotes}
\usepackage{bigints}
\usepackage{amsmath}
\usepackage{xcolor}
\usepackage{booktabs}
\usepackage{relsize}
\usepackage{float}
\usepackage[section]{placeins}
\definecolor{dgreen}{RGB}{0,204,0}
\definecolor{dora}{RGB}{255,150,0}
\definecolor{red1}{RGB}{255,0,0}

\usepackage{orcidlink}

\newcommand{\be}{\begin{equation}}
\newcommand{\ee}{\end{equation}}
\newcommand{\bea}{\begin{eqnarray}}
\newcommand{\eea}{\end{eqnarray}}

\newcommand{\nonu}{\nonumber}
\renewcommand{\v}{\mathrm{v}}

\begin{document}

\title{Massive cold hybrid stars in a modified Polyakov-Nambu-Jona-Lasinio model}

\author{Sk Md Adil Imam$^{1,2}$\orcidlink{0000-0003-3308-2615}}
\email{adil.imam@unab.cl}
\author{Pedro Costa$^2$\orcidlink{0000-0003-4809-6542}}
\email{pcosta@uc.pt}
\author{Mariana Dutra$^3$\orcidlink{0000-0001-7501-0404}}
\email{marianad@ita.br}
\author{Odilon Lourenço$^3$\orcidlink{0000-0002-0935-8565}}
\email{odilon@ita.br}
\author{Renan Pereira$^2$\orcidlink{0000-0003-3794-7719}}
\author{Constança Provid\^encia$^2$\orcidlink{0000-0001-6464-8023}}
\email{cp@uc.pt}

\affiliation{
$^1$Institute of Astrophysics, Department of Physics and Astronomy, Universidad Andrés Bello, Santiago, Chile
\\
$^2$CFisUC, Department of Physics, University of Coimbra, P-3004 - 516 Coimbra, Portugal
\\
$^3$Departamento de Física e Laboratório de Computação Científica Avançada e Modelamento (Lab-CCAM), Instituto Tecnológico de Aeronáutica, DCTA, 12228-900, São José dos Campos, SP, Brazil
}
\date{\today}

\begin{abstract}

We propose a modified Polyakov-loop Nambu--Jona-Lasinio (mPNJL) model in which the Polyakov potential is given by an explicit dependence on the quark chemical potential, allowing it to remain finite at zero temperature and thus to describe the confinement-deconfinement transition in cold dense matter. Combining this modified quark sector with hadronic equations of state via a Maxwell construction, we find that, depending on the model parameters, the equation of state can exhibit either two phase transitions, from hadronic matter to confined quark matter and subsequently to deconfined quark matter, or a single transition directly from hadronic to deconfined quark matter or from  hadronic to confined quark matter. Stable massive cold hybrid stars with only confined and/or deconfined quark phase are obtained. We systematically examine how the parameters of the modified Polyakov potential and the quark vector interactions control the location of these transitions, and find that repulsive vector interactions are essential to obtain a stable quark core. Hybrid stars with confined and/or a deconfined core can reach maximum masses above $2M_\odot$, provided a sufficiently stiff hadronic equation of state is used at low density. In the core of the maximum-mass configurations, the speed of sound shows variations at finite baryon densities, with c$_s^2(\mu)$ departing from the asymptotic conformal value c$_s^2$ = 1/3 in confined core stars. These variations serve as a diagnostic of the equation-of-state stiffness while remaining fully consistent with causality and thermodynamic stability. This work establishes the qualitative role of each model parameter in shaping hybrid-star structure.
\end{abstract}

\maketitle

\section{Introduction}
The study of matter under extreme conditions of density and temperature, which are reached in relativistic heavy-ion collisions or in the interior of compact stars, remains one of the most important challenges in modern nuclear physics, particle physics, and astrophysics. Under such conditions, conventional descriptions of matter based solely on nucleonic degrees of freedom may become inadequate, and the emergence of exotic components such as hyperons, meson condensates,  deconfined quark matter,  or a  quark-gluon plasma (QGP)  becomes increasingly plausible~\cite{ Bonanno:2011ch, Benic:2014jia, Miyatsu:2015kwa,Zhao:2020dvu,Malik:2022jqc}. 

Recent advances in both terrestrial experiments and multimessenger astrophysics have provided unprecedented opportunities to probe the properties of strongly interacting matter under extreme conditions. Relativistic heavy-ion collision programs at RHIC~\cite{RHIC} and the LHC~\cite{LHC_CERN} have produced strongly interacting matter at high temperatures and low baryonic densities, offering valuable insights into the quark--gluon plasma and the Quantum Chromodynamics (QCD) phase diagram. Future facilities, including the forthcoming Electron Ion Collider~\cite{EIC} at Brookhaven, FAIR~\cite{FAIR} experiment at GSI and the NICA complex at JINR~\cite{NICA}, are specifically designed to explore the high-baryonic-density region where the onset of deconfinement and the possible existence of a critical endpoint may occur. 

Complementary information is obtained from astrophysical observations of neutron stars. Precise mass measurements of massive  pulsars, such as PSR J1614-2230~\cite{Demorest:2010bx,NANOGrav:2019jur}, PSR J0348+0432~\cite{Antoniadis:2013pzd}, or PSR J0740+6620~\cite{Fonseca:2021wxt}, together with the mass-radius constraints from the NICER mission observations, PSR J0437+4715~\cite{Choudhury:2024xbk}, PSR J0614+3329~\cite{Mauviard:2025dmd}, PSR J0030+0451~\cite{Riley:2019yda, Miller:2019cac, Vinciguerra_2024}, and PSR J0740+6620~\cite{Riley:2021pdl, Miller:2021qha, Salmi_2024, Dittmann_2024}, as well as tidal deformability measurements from gravitational-wave events like GW170817~\cite{Abbott:2017vwq, Abbott:2018exr, Abbot2019}, have imposed stringent constraints on the equation of state of dense matter. Future detectors, including the Einstein Telescope~\cite{Punturo:2010zz} and Cosmic Explorer~\cite{Reitze:2019iox}, are expected to sharpen these constraints considerably. These developments establish a strong connection between laboratory experiments, astrophysical observations, and theoretical studies of the QCD phase structure.

The behavior of strongly interacting quarks and gluons is described by QCD, the non-Abelian gauge theory underlying the strong nuclear force~\cite{Weinberg_1973,Fritzsch:1973pi}. The validity of perturbative QCD is reduced at finite baryonic density due to the enhancement of the running strong coupling~\cite{Gross_1973,Alford_2008}. However, results from perturbative QCD at very high densities~\cite{Kurkela:2009gj} impose constraints on the equation of state of baryonic matter at densities existing inside neutron stars~\cite{Komoltsev:2021jzg}. Although lattice QCD predicts a crossover transition between hadronic and quark–gluon plasma phases at high temperature and vanishing chemical potential, calculations at finite baryonic density are still limited by the well-known fermion sign problem~\cite{Splittorff_2007}. The possibility of a first-order transition line terminating at a critical endpoint has been predicted in a broad range of effective models, e.g. MIT bag~\cite{DeGrand_1975} and the Nambu-
Jona-Lasinio (NJL)~\cite{Nambu_1961,Nambu2_1961,BUBALLA_2005,Vogl:1991qt,Klevansky_1962}.

In the NJL model, constituent quark masses are generated dynamically through local four-fermion interactions. Although the model successfully describes spontaneous chiral symmetry breaking, it lacks explicit gluonic degrees of freedom and therefore cannot account for color confinement or the confinement–deconfinement transition~\cite{Gross_1973}. The confinement deficiency of the NJL model is partially remedied in the Polyakov NJL~(PNJL) framework by coupling quarks to a background temporal gluon field represented by the Polyakov loop~\cite{Fukushima:2003fw,Ratti_2007,Ratti2_2007,BRATOVIC_2013,Claudia_2006,Fukushima_2008}. The traced Polyakov loop, $\Phi$, acts as an approximate order parameter for deconfinement and effectively incorporates gluonic dynamics, although it yields only statistical rather than true color confinement. A limitation of the standard PNJL framework is that the Polyakov loop potential vanishes in the zero-temperature limit, the condition encountered in cold neutron and hybrid stars~(HSs). To overcome this limitation and investigate the confinement--deconfinement transition in cold neutron and hybrid stars, we employ a modified Polyakov loop potential that remains finite in the zero-temperature limit and is dependent on the quark chemical potentials. A similar approach has been employed in~\cite{Dexheimer_2010} in a hadronic SU(3) nonlinear $\sigma$ model with quark degrees of freedom included. Other versions of the PNJL model were also proposed for this goal, with the Polyakov loop potential depending explicitly on the quark densities and quark condensates~\cite{Mattos_2019,Mattos:2021alf,Mattos2_2021}, and on the baryonic density~\cite{Ivanytskyi_2019}.

Within this description, we  build hybrid star equations of state with two phase transitions, from hadron to quark confined matter and from confined to deconfined quark matter, or with just one phase transition, directly from hadron to deconfined quark matter. It is shown that it is possible to describe two solar mass hybrid stars if a vector term is included in the model. The primary purpose of the present work is to establish the modified PNJL framework and to identify the role played by each model parameter in determining the phase structure and hybrid-star properties.

The paper is organized as follows. In Sec.~\ref{Methe}, we describe the formalism of the modified Polyakov-Nambu-Jona-Lasinio model~(mPNJL) model and the thermodynamics of asymmetric quark matter. Sec.~\ref{Res} presents the results for hybrid stars and their dependence on the model parameters. Finally, Sec.~\ref{Conc} summarizes our conclusions. Throughout this paper, we adopt natural units, namely, $G=c=\hbar=1$.

\section{Formalism}
\label{Methe}

\subsection{PNJL model}

For a three-flavor quark system with vector interactions included, the Lagrangian density of the conventional NJL model can be given by~\cite{Mishustin_2000,Ferreira:2020kvu} 
\begin{align}
\mathcal{L}_{\rm{NJL}} &=\bar q(i\gamma_\mu \partial^\mu-\hat m )q 
\nonumber\\
&+\frac{G_S}{2}\sum_{a=0}^{8}
\Big[
(\bar q \lambda_a q)^2
-(\bar q i\gamma_5 \lambda_a q)^2
\Big]
\nonumber\\
&-\frac{G_V}{2} \sum_{a=0}^{8}
\Big[
(\bar q \gamma^\mu \lambda_a q)^2
+(\bar q \gamma^\mu \gamma_5 \lambda_a q)^2
\Big]
\nonumber\\
&+K \Big[
{\rm det}_f \bigl(\bar q (1-\gamma_5) q\bigr) +{\rm det}_f \bigl(\bar q (1+\gamma_5) q\bigr)
\Big]\nonumber\\
& - G_{\v\v}\left[\left(\bar{\psi}\gamma^\mu\lambda_0\psi\right)^2
+\left(\bar{\psi}\gamma^\mu\gamma_5\lambda_0\psi\right)^2\right]^2.
\label{Eq:NJL}
\end{align}
In this expression, \(q\equiv(q_u,q_d,q_s)^T\) denotes the quark field vector in flavor space, while \(\hat{m}=\mathrm{diag}(m_u,m_d,m_s)\) represents the current quark mass matrix. The matrices \(\lambda_a\) ($a=1,...,8)$ correspond to the generators of the SU(3) flavor group, and $\lambda_0$ is proportional to the identity matrix. The determinant term in flavor 
space describes the six-fermion ’t Hooft interaction~\cite{Hooft_76}, which explicitly breaks the axial \(U_A(1)\) symmetry~\cite{Klimt_90} and accounts for the anomalously large \(\eta^\prime\) meson mass. This interaction was originally introduced by Kobayashi and Maskawa~\cite{Kobayashi_70}. The PNJL model generalizes Eq.~\eqref{Eq:NJL} to~\cite{Fukushima:2003fw,Ratti_2007,Ratti2_2007,BRATOVIC_2013,Claudia_2006,Fukushima_2008}
\begin{align}
 \mathcal{L}_{\rm{PNJL}} = \mathcal{L}_{\rm{NJL}} - \bar q\gamma_\mu A^\mu q - \mathcal{U}(\Phi,\bar{\Phi},T),
\label{Eq:PNJL}
\end{align}
 i.e., it incorporates coupling between quarks and a static background gluon field through the covariant derivative $D^\mu\equiv \partial^\mu+iA^\mu$, where $A_\mu=\delta_\mu^0 A_0$ and $A_0=gA_0^a \lambda_a/2$ with $g$ denoting the gauge coupling constant and \(A_0\) the gluonic background field. In addition, the model includes the effective Polyakov loop potential \(\mathcal{U}(\Phi,\bar{\Phi},T)\), which encodes essential aspects of color confinement. The effective Polyakov loop potential depends on the traced Polyakov loop \(\Phi\) and its complex conjugate \(\bar{\Phi}\). 

The Polyakov loop $\Phi$ serves as the order parameter for the color deconfinement transition. 
It is constructed from the temporal gauge field component $A_4$, integrated over the imaginary 
time interval. The loop characterizes confinement: $\Phi \approx 0$ indicates the confined phase, 
while $\Phi > 0$ signals deconfinement.

\begin{equation}
\Phi = \frac{1}{3} \mathrm{Tr} \exp \left( i \int_0^{1/T} d\tau \, A_4 \right) \quad
\end{equation}

In effective QCD approaches, the confinement--deconfinement transition can be characterized through the Polyakov loop, defined as a Wilson line in Euclidean space-time. Its normalized trace, $\Phi$, is related to the free energy of a static quark $F_q$ via $\Phi \sim e^{-F_q/T}$. In the confined phase $F_q \rightarrow \infty$ and $\Phi \rightarrow 0$, while in the deconfined phase $F_q \rightarrow 0$ and $\Phi \rightarrow 1$. Thus, $\Phi$ serves as an order parameter for deconfinement in the pure gauge sector and is linked to the realization of the center symmetry $Z(3)$\cite{SvetitskyYaffe1982}.   In our mean-field effective model, the Polyakov loop $\Phi$ provides an approximate 
order parameter for deconfinement, capturing effective confinement dynamics relevant 
to cold dense matter rather than exact QCD confinement. The physical validity of our 
results rests on consistency checks---stellar stability, causality, and thermodynamic 
stability---demonstrating the robustness of the approach. Within the mean-field approximation, and using $
\Phi=\bar{\Phi}$, see Refs.~\cite{Ratti_2007,ROBNER_2008,DEXHEIMER_2009,Steinheimer_2010,BRATOVIC_2013,Ferreira:2018sun,Mattos_2019,Hansen:2019lnf,Mattos:2021alf,Dexheimer_21}, Eq.~\eqref{Eq:PNJL} becomes
\begin{align}
&\mathcal{L}_{\mathrm{PNJL}}^{\mathrm{MFA}} =
\sum_f \bar{q}_f\left(i\gamma^\mu D_\mu - M_f -\gamma^0 W_f \right)q_f - G_S \sum_f \rho_{sf}^2
\nonumber\\
&+G_V \sum_f \rho_{\v f}^2 -4K \prod_f \rho_{sf}-\mathcal{U}(\Phi,T)
+ \frac{4}{3}G_{\v\v}(\sum_f \rho_{\v f})^4,
\label{eq:PNJL_MFA}
\end{align}
with
\begin{align}
W_f = 2G_V\rho_{\v f} + \frac{16}{9}G_{\v\v}(\sum_{j=u,d,s} \rho_{\v j}\,\,)^3,
\label{eq:Wf}
\end{align}
where $\rho_{sf}=\langle \bar{q}_f q_f\rangle$ and $\rho_{\v f}=\langle q_f^\dagger q_f\rangle$ denote the scalar and vector densities, respectively. The degeneracy factor is given by $\gamma=N_s\times N_c=6$, where \(N_s=2\) and \(N_c=3\) correspond to the spin and color degrees of freedom, respectively. The constituent quark masses, dynamically generated through the quark condensates are respectively given by
\begin{equation}
M_f = m_f - 2G_S \rho_{sf} - 2K\prod_{f'\neq f}\rho_{sf'}.
\label{eq:constituent_mass}
\end{equation}

The corresponding grand-canonical thermodynamic potential density reads
\begin{align}
&\Omega_{\mathrm{PNJL}} = G_S \sum_f \rho_{sf}^2 - G_V \sum_f \rho_{\v f}^2 + 4K \prod_f \rho_{sf}\nonumber\\
&-\frac{4}{3}G_{\v\v}(\sum_f\rho_{\v f})^4
\nonumber\\
&- \frac{\gamma}{6\pi^2}\sum_f\int_0^\infty\frac{dk\, k^4}{E_f}\left[f(E_f,T,\Phi)
+\bar{f}(E_f,T,\Phi)\right]
\nonumber\\
&- \frac{\gamma}{2\pi^2}\sum_f \int_0^\Lambda dk\, k^2 E_f  +\mathcal{U}(\Phi,T),
\label{eq:Omega_PNJL}
\end{align}
where \(\gamma\) is the spin-color degeneracy factor, as already mentioned, and $E_f=(k^2 + M_f^2)^{1/2}$ is the single-particle energy for quarks of flavor \(f\). The generalized Fermi-Dirac distributions for quarks and antiquarks are denoted by $f(E_f,T,\Phi)$ and $\bar{f}(E_f,T,\Phi)$, see Refs.~\cite{Fukushima_2008,Claudia_2006,Ratti2_2007,Mattos2_2021} for the respective analytical expressions.

A commonly used form of the Polyakov loop potential~\cite{Claudia_2006,Ratti2_2007,Ratti_2007,Fukushima_2008,DEXHEIMER_2009,Dexheimer_2010,Costa_2010} is given by 
\begin{align}
\frac{\mathcal{U}(\Phi,T)}{T^4} = -\frac{a(T)}{2}\,\Phi^2 + b(T)\ln\left[X(\Phi)\right],
\label{Eq:PL}
\end{align}
with $X(\Phi)=1-6\Phi^2+8\Phi^3-3\Phi^4$ and the temperature-dependent coefficients 
\begin{align}
a(T)&=a_0+a_1\left(\frac{T_0}{T}\right) + a_2\left(\frac{T_0}{T}\right)^2,
\\
b(T)&=b_3\left(\frac{T_0}{T}\right)^3.
\nonumber
\end{align}
The parameters $T_0$, $a_0$, $a_1$, $a_2$, and $b_3$ are fixed by fitting pure gauge lattice QCD results~\cite{Ejiri_1998,Boyd_1996,Kaczmarek_2002}. In Ref.~\cite{Ratti_2007}, for instance, the parametrization used is $T_0 = 270$~MeV, with
\begin{align}
a_0 = 3.51\quad
a_1 = -2.47\quad
a_2 = 15.2\quad
b_3 = -1.75.
\label{param_pnjl}
\end{align}

In the presence of dynamical quarks, $T_0$ may depend on the number of flavors and can also vary with the quark chemical potentials~\cite{Fischer_2014,Schaefer_2007,Abuki_2008,Shao_2016}. We are going to use $T_0$ as a free parameter. A well-known limitation of the standard PNJL model is that the Polyakov loop potential vanishes as $T \rightarrow 0$, effectively reducing the model to the NJL framework. Consequently, the deconfinement order parameter becomes irrelevant in the cold and dense regime, preventing a consistent description of quark or hybrid stars. To describe such compact objects within a framework that incorporates an order parameter for the confinement--deconfinement phase transition, it is necessary to employ a modified version of the PNJL model. In particular, the effective Polyakov loop potential must be reformulated so that it remains finite at \(T=0\). This will be discussed in the next subsection. 
\\ 
\subsection{Formulation of the modified PNJL model}
\label{mPNJL}
  
Following the approach of Ref.~\cite{Xin_2014}, we reformulate the effective Polyakov-loop potential by modifying both sides of Eq.~\eqref{Eq:PL}. The left-hand side is based on the QCD equation of state in the Stefan–Boltzmann limit, corresponding to an ideal gas of quarks and gluons at leading order in the coupling constant. The pressure for a gas of massless quarks $P_q$ reads
\begin{align}
P_q = \sum_f P_f = N_c N_f \frac{7\pi^2 T^4}{180} + N_c \sum_f \left(\frac{T^2\mu_f^2}{6}
+ \frac{\mu_f^4}{12\pi^2}\right),
\end{align}
and the pressure for a gas of massless gluons is
\begin{equation}
P_G = 2\left(N_c^2-1\right)\frac{\pi^2 T^4}{90}.
\end{equation}
The Stefan-Boltzmann pressure, given by the sum of these two contributions, is
\begin{align}
P_{\rm SB} &= P_q + P_G = \frac{19\pi^2 T^4}{36} + \sum_f\left(\frac{T^2\mu_f^2}{2}+\frac{\mu_f^4}{4\pi^2}\right) 
\nonumber\\
&= \frac{19\pi^2}{36}\left[T^4+\sum_f\left(\frac{18}{19\pi^2}T^2\mu_f^2+
\frac{9}{19\pi^4}\mu_f^4\right)
\right].
\end{align}

The first modification to the effective Polyakov loop potential is the substitution of the global $T^4$ dependence on the left side of Eq.~\eqref{Eq:PL} with the dependence in the Stefan--Boltzmann pressure, namely,
\begin{equation}
T^4 \rightarrow T^4 + \sum_f\left(\frac{18}{19\pi^2}T^2\mu_f^2 + \frac{9}{19\pi^4}\mu_f^4\right).
\end{equation}
We emphasize that the Stefan-Boltzmann pressure serves as a dimensional and 
thermodynamic guide for extending the functional form of the Polyakov potential 
from the finite-temperature regime to finite chemical potential. It provides the 
correct scaling and structure needed to capture confinement dynamics at neutron 
star densities, rather than a quantitatively accurate description of cold strongly 
interacting matter.

Note that the standard effective Polyakov loop potential is recovered in the limit $\mu_f = 0$. This modification is motivated by the Dyson--Schwinger analysis presented in Ref.~\cite{Fischer_2014}. The chemical potential dependence on the right side of Eq.~\eqref{Eq:PL} will be given by $\mu_f$ dependent $T_0$. Perturbative calculations (hard thermal loop and hard dense loop calculations of perturbative QCD) allow the following
substitution~\cite{Schaefer_2007}
\begin{align}
\frac{T_0}{T} \;\rightarrow\; \frac{T_0(\mu_f)}{T}
= \frac{T_\tau}{T}\, \exp\!\left(-\frac{1}{C_1 - C_2 \mu_f^2}\right).
\label{Eq:T0_T}
\end{align}
Here, the constants $C_1$, and $C_2$ are 
\begin{align}
C_1
&=
\frac{\alpha_0}
{6\pi}(11N_c-2N_f),
\label{eq:C1}
\\[2mm]
C_2
&=
\frac{16\alpha_0}
{\pi}\,
\frac{N_f}{T_\tau^2}.
\label{eq:C2}
\end{align}
The parameters \(T_0\) and \(\alpha_0\) are free constants to be fixed.

The ratio $T_0(\mu_f)/T$ in Eq.~\eqref{Eq:T0_T} diverges as $T \to 0$, precisely in the regime of interest. Following Refs.~\cite{Xin_2014,Pereira2016}, we regularize the low-temperature behavior of the modified Polyakov loop potential by introducing the phenomenological replacement
\begin{equation}
\frac{T_0}{T} \;\to\; \frac{T_0}{\sqrt{T^2 + g(\mu_f)}} ,
\label{eq:polyakov_regularization}
\end{equation}
where $g(\mu_f)$ is written as a power series in $\mu_f$ given by
\begin{equation}
g(\mu_f) = \sum_{n=1}^{\infty} \eta_n \mu_f^{n},
\label{Eq:g_muf}
\end{equation}
where the summation is chosen to start at $n=1$ in order to recover the standard expression $T_0/T$ in the limit $\mu_f=0$. The coefficients $\eta_i$ are determined by Taylor expanding both Eqs.~\eqref{Eq:T0_T} and~\eqref{eq:polyakov_regularization} in powers of $\mu_f$ around $\mu_f=0$, evaluating the expansions at the lattice-QCD deconfinement temperature $T=T_{\mathrm{dec}}^{\mathrm{lat}}$, and matching the coefficients order by order. Up to second order in $\mu_f$, this procedure leads to
\begin{align}
\quad T_0 &= T_\tau e^{-1/C_1}, 
\end{align}
from the zeroth order in $\mu_f$, and 
\begin{align}
&\eta_2 = 2\left(T_{\mathrm{dec}}^{\mathrm{lat}}\right)^2 \frac{C_2}{C_1^2} 
\nonumber\\
&= \frac{1152\pi N_f(T_{\mathrm{dec}}^{\mathrm{lat}})^2}{\alpha_0(11N_c-2N_f)^2T^2_0}\exp\left[-\frac{12\pi}{\alpha_0(11N_c-2N_f)}
\right],
\label{Eq:eta2}
\end{align}
from the second order in $\mu_f$. Since $N_f=N_c=3$, and using $T_{\mathrm{dec}}^{\mathrm{lat}}=170$~MeV~\cite{Gupta_Science}, one has $\eta_2=\eta_2(T_0,\alpha_0)$. In the next section, we will show the effect of these two specific parameters, $T_0$ and $\alpha_0$, in the mPNJL model.

Notice that, from this procedure, the functions $a(T)$ and $b(T)$ acquire an explicit dependence on the chemical potential, becoming $a(T,\mu_f)$ and $b(T,\mu_f)$. Since different quark flavors may contribute differently, particularly when their chemical potentials are not identical (as is the case for $\beta-$equilibrated matter, see Sec.~\ref{Thermo}), it becomes necessary to express these quantities as sums over all quark flavors:
\begin{align}
a(T)&\rightarrow \frac{1}{N_f}\sum_f a(T,\mu_f),
\\
b(T)&\rightarrow \frac{1}{N_f}\sum_f b(T,\mu_f).
\end{align}
The factor $1/N_f$ is a normalization constant that allows us to recover the standard Polyakov loop effective potential, Eq.~\eqref{Eq:PL}, in the limit $\mu_f \to 0$.

By applying these two changes, the modified Polyakov loop potential can finally be written as
\begin{align}
&\frac{\mathcal{U}(\Phi,T,\mu_f)}{T^4+\displaystyle\sum_f \left(\frac{18}{19\pi^2}T^2\mu_f^2
+ \frac{9}{19\pi^4}\mu_f^4\right)} = 
\nonumber\\
&-\frac{\Phi^2}{2N_f}\sum_f a(T,\mu_f) + \frac{1}{N_f}\sum_f b(T,\mu_f)\ln\left[X(\Phi)\right],
\label{UmPNJL}
\end{align}
 where $a(T,\mu_f)$ and $b(T,\mu_f)$ are now functions that depend on temperature and quark chemical potential in the following way
\begin{align}
a(T,\mu_f)&=a_0+a_1\frac{T_0}{\sqrt{T^2+g(\mu_f)}} + a_2\frac{T_0^2}{T^2+g(\mu_f)},
\\
b(T,\mu_f)&=b_3\frac{T_0^3}{\left[T^2+g(\mu_f)\right]^{3/2}},
\end{align}
with
\begin{align}
g(\mu_f) = \eta_2(T_0,\alpha_0)\mu_f^2.
\end{align}
Therefore, the Lagrangian density in the mean-field approximation of the mPNJL model is
\begin{align}
\mathcal{L}_{\mathrm{mPNJL}}^{\mathrm{MFA}} &=
\sum_f \bar{q}_f\left(i\gamma^\mu D_\mu - M_f -\gamma^0 W_f \right)q_f - G_S \sum_f \rho_{sf}^2
\nonumber\\
&+G_V \sum_f \rho_{\v f}^2 -4K \prod_f \rho_{sf}-\mathcal{U}(\Phi,\mu_f,T) 
\nonumber \\ 
&+ \frac{4}{3}G_{\v\v}(\sum_f \rho_{\v f})^4,
\label{eq:mPNJL}
\end{align}
with the Polyakov loop potential given in Eq.~\eqref{UmPNJL}. From this quantity, one obtains the pressure of the system. At $T=0$ it reads
\begin{align}
&P_{\mathrm{mPNJL}} = - G_S \sum_f \rho_{sf}^2 + G_V \sum_f \rho_{\v f}^2 
\nonumber\\
&- 4K \prod_f \rho_{sf} +\frac{4}{3}G_{\v\v}(\sum_f\rho_{\v f})^4 \nonu \\
&+ \frac{\gamma}{2\pi^2} \sum_f \int_0^{\Lambda} dk\, k^2 (k^2 + M_f^2)^{1/2} + \Omega_{\mathrm{vac}}
\nonu \\
&+ \frac{\gamma}{6\pi^2} \sum_f \int_0^{k_{F_f}} dk\, \frac{k^4}{(k^2 + M_f^2)^{1/2}} - \mathcal{U}(\Phi,\mu_f),
\label{eq:P_T0}
\end{align}
with $\mathcal{U}(\Phi,\mu_f)\equiv \mathcal{U}(\Phi,T=0,\mu_f)$, and the vector densities given by $\rho_{\v f} = \gamma k_{F_f}^3/(6\pi^2)$ where $k_{F_f}$ is the Fermi momentum of the quark~$f$. The constant term $\Omega_{\mathrm{vac}} = -P_{\mathrm{vac}}$ is included to guarantee that, in the vacuum limit ($\rho_u = \rho_d = \rho_s = 0$), the thermodynamic potential satisfies $\Omega_{\mathrm{mPNJL}}(\rho_f=0) = -P_{\mathrm{mPNJL}}(\rho_f=0)=0$. The quark chemical potentials are given by
\begin{equation}
\mu_f = (k_{F_f}^2 + M_f^2)^{1/2} + W_f.
\label{eq:muf}
\end{equation} 

The inclusion of an explicit chemical potential dependence in the Polyakov loop potential leads to corresponding modifications in the quark densities as follows
\begin{align}
&\rho_f = - \frac{\partial \Omega_\mathrm{mPNJL}}{\partial \mu_f} = \rho_{\v f} - \frac{\partial \mathcal{U}(\Phi,\mu_f)}{\partial \mu_f} = \rho_{\v f} 
\nonumber\\
&+ \frac{9}{19\pi^4 N_f}\Bigg[2\mu_f^3 \Phi^2 \sum_{j=u,d,s}a(\mu_j) 
+ \frac{\Phi^2}{2} \frac{d a(\mu_f)}{d\mu_f}\sum_{j=u,d,s} \mu_j^4
\nonumber\\
&- \ln X(\Phi)\Bigg(4\mu_f^3 \sum_{j=u,d,s} b(\mu_j) + \frac{d b(\mu_f)}{d\mu_f} \sum_{j=u,d,s} \mu_j^4\Bigg)\Bigg],
\label{eq:densities}
\end{align}
with $a(\mu_f)\equiv a(T=0,\mu_f)$ and $b(\mu_f)\equiv b(T=0,\mu_f)$. The baryonic density is defined as $\rho_B=(\rho_u+\rho_d+\rho_s)/3$. Finally, the energy density is obtained through the Euler relation as
\begin{align}
&\mathcal{E}_{\mathrm{mPNJL}} = -P_{\mathrm{mPNJL}} + \sum_f\mu_f\rho_f
\nonumber\\
&=G_S \sum_f \rho_{sf}^2 + G_V \sum_f \rho_{\v f}^2 + 4K \prod_f \rho_{sf} \nonumber\\
& + \frac{4}{9}G_{\v\v}(\sum_f\rho_{\v f})^4
-\sum_f\mu_f\frac{\partial\mathcal{U}}{\partial\mu_f} - \Omega_{\mathrm{vac}}
\nonu \\
&- \frac{\gamma}{2\pi^2} \sum_f \int_{k_{F_f}}^{\Lambda}dk\, k^2 (k^2 + M_f^2)^{1/2} 
+ \mathcal{U}(\Phi,\mu_f).
\end{align}
The functional forms for the quark constituent masses are given in Eqs.~\eqref{eq:constituent_mass}. The analytical expressions for the quark condensates are
\begin{equation}
\rho_{sf} = - \frac{\gamma M_f}{2\pi^2} \int_{k_{F_f}}^{\Lambda} dk \,  \frac{k^2}{(k^2 + M_f^2)^{1/2}}.
\label{Eq:quark_cond}
\end{equation}
Notice that in the previous expressions an usual regularization only applies a cutoff (parameter $\Lambda$) to the divergent integrals.
 
An important advantage of the proposed potential $\mathcal{U}(\Phi,\mu_f)$ is that it admits an analytical solution for the Polyakov loop~\cite{Pereira2016}. It is obtained by imposing the stationary condition $\partial\Omega/\partial\Phi=0$, which leads to
\begin{equation}
(\Phi - 1)^2\Phi\left(A + 12B + 2A\Phi - 3A\Phi^2 \right) = 0,
\label{Eq:PL_sol}
\end{equation}
where $A=\sum_f a(\mu_f)$, and $B=\sum_f b(\mu_f)$. The solutions of this equation are
\begin{align}
\Phi &= 0,
\label{Eq:PL_sol1}\\
\Phi &= 1 \quad (\text{double solution}),
\label{Eq:PL_sol2}\\
\Phi^{\mp} &= \frac{1}{3} \mp \frac{2}{3}\sqrt{1+\frac{9B}{A}}.
\label{Eq:PL_sol3}
\end{align}

The first solution, Eq.~(\ref{Eq:PL_sol1}), corresponds to a vanishing Polyakov loop and is associated with the confined phase. The second solution, Eq.~(\ref{Eq:PL_sol2}), represents the limiting value of the Polyakov loop, $\Phi=1$. 
Finally, the solutions given by Eq.~(\ref{Eq:PL_sol3}) describe the deconfined regime with dynamical quarks included, in which the Polyakov loop assumes values in the interval $0<\Phi<1$. In our model, the onset of deconfinement is identified by the condition $\Phi>0$. Therefore, these solutions characterize the deconfined phase of asymmetric quark matter in $\beta$ equilibrium with leptons, which is the scenario investigated here, see Sec.~\ref{Thermo}.
\subsection{mPNJL model in $\beta-$equilibrated matter}
\label{Thermo}

In the present study, all calculations are performed in the zero-temperature limit. At $T = 0$, the regularization scheme in Eq.~\eqref{eq:P_T0} with cutoff $\Lambda$ is consistent with Eq.~\eqref{eq:Omega_PNJL}, as the medium contribution reduces to an integral over the occupied Fermi sea with an effective cutoff at the Fermi momentum. Having established the fundamental equations governing the SU(3) mPNJL framework, we can now investigate its thermodynamic behavior. To proceed with the numerical analysis, it is necessary to specify the model parameters, which include the coupling constants $G_S$, $G_V$, $G_{\v\v}$, and $K$, the quark masses $m_u$, $m_d$, and $m_s$, as well as the momentum cutoff~$\Lambda$. For this study we use a modified version of the HK parameter set~\cite{Hatsuda_1994}: $\Lambda = 631.4$~MeV, $G_S\Lambda^2 = 3.562$, $K\Lambda^5 = -9.29$, $m_{u,d} = 5.5$~MeV and $m_s = 135.7$~MeV. This modification is implemented so that the constituent $u$ and $d$ quark masses in vacuum are approximately equal to one-third of the nucleon mass. Along with accurately reproducing the nucleon mass in vacuum, the chosen parametrization yields a reasonable agreement with the observed vacuum properties of the mesons, as one can see in Table~\ref{tab:meson_properties}.
\begin{table}[!htb]
\centering
\caption{Comparison between the SU(3) modified HK parametrization and experimental values~\cite{Olive_2014}.}
\label{tab:meson_properties}
\begin{tabular}{lcc}
\toprule
Observable & SU(3) parameterization & Experimental~\cite{Olive_2014} \\
\midrule
$m_{\pi}$ [MeV]     & 138.2 & 139.6 \\
$f_{\pi}$ [MeV]     & 90.9  & 92.2  \\
$m_{K}$ [MeV]       & 493.9 & 493.7 \\
$f_{K}$ [MeV]       & 96.4  & 110.4 \\
$m_{\eta}$ [MeV]    & 479.6 & 547.9 \\
\bottomrule
\end{tabular}
\label{tab}
\end{table}
We also use the parametrization given in Eq.~\eqref{param_pnjl} for the constants of the Polyakov loop potential of the model.

An important source of uncertainty in effective quark models is the determination of the vector interaction strength. Although theoretical considerations based on the underlying quark interaction suggest a specific relation between the vector and scalar couplings, the effective value of $G_V$ remains model dependent and is not uniquely established. For this reason, different studies have adopted distinct prescriptions for its determination. Phenomenological analyses often allow the vector coupling to vary in order to improve the description of hadronic observables or to account for modifications induced by the surrounding medium~\cite{Schafer_1995}. Since the properties of strongly interacting matter are expected to change with temperature and density, the effective strength of the vector channel may differ from its vacuum estimate. The choice of $G_V$ is particularly relevant when investigating the thermodynamics of dense matter. Variations in this parameter can significantly alter the predicted phase structure, affecting both the nature and the location of phase transitions~\cite{BRATOVIC_2013,Hell_2013,Denke:2013gha,Costa:2015bza}. In many cases, a stronger repulsive vector interaction tends to weaken first-order transitions and can even eliminate them from the phase diagram. Owing to its substantial influence on the model predictions, we do not impose a fixed value for $G_V$. Instead, as in~\cite{CamaraPereira:2016chj,Ferreira:2020kvu,Mattos:2021alf,Mattos2_2021}, it is treated as a free parameter, allowing us to systematically examine its impact on the physical quantities studied in this work.

The eight-quark vector-isoscalar coupling is subject to the same source of uncertainty: it is likewise not fixed uniquely, and its effective strength, $G_{\v\v}$, is expected to be similarly medium dependent~\cite{BUBALLA_2005,Fukushima_2008}. However, $G_V$ and $G_{\v\v}$ affect the equation of state in qualitatively different ways. While $G_V$ enters the effective quark chemical potential linearly in the density, $G_{\v\v}$ enters cubically~\cite{Ferreira:2020kvu}, so its contribution is negligible at the onset of quark matter and only becomes significant once the density is already large. As a result, $G_V$ mainly controls where the confinement--deconfinement transition sets in, whereas $G_{\v\v}$ mainly controls how stiff the quark equation of state becomes once quark matter is realized, with a correspondingly stronger influence on the speed of sound at the stellar center, on the mass and radius of the maximum-mass configuration, and on the extent of the quark core, than $G_V$ has~\cite{Ferreira:2020kvu}. Owing to this substantial and distinct influence on the model predictions, we likewise do not impose a fixed value for $G_{\v\v}$. Instead, as in ~\cite{CamaraPereira:2016chj,Ferreira:2020kvu}, it is treated as a free parameter, allowing us to systematically examine its impact, alongside $G_V$, on the location of the confined-to-deconfined transition and on the resulting hybrid-star
structure.

The system under consideration is assumed to be in chemical equilibrium under weak interactions and to satisfy electric charge neutrality. The equilibrium conditions impose the following relations among the chemical potentials of quarks and leptons:
\begin{equation}
\mu_d=\mu_s=\mu_u+\mu_e,
\qquad
\mu_\mu=\mu_e.
\end{equation}
Introducing a common quark chemical potential $\mu$, related to the baryonic chemical potential through $\mu_B=3\mu$, the individual quark chemical potentials can be expressed as
\begin{equation}
\mu_u=\mu-\frac{2}{3}\mu_e,
\qquad
\mu_d=\mu_s=\mu+\frac{1}{3}\mu_e.
\label{eq:quark_chemical_potentials}
\end{equation}
The condition of electric charge neutrality requires the total positive and negative charge densities to balance, yielding
\begin{equation}
\frac{2}{3}\rho_u-\frac{1}{3}\rho_d-\frac{1}{3}\rho_s
=\rho_e+\rho_\mu.
\label{eq:charge_neutrality}
\end{equation}
For a free relativistic electron gas, the electron number density (for massless electrons) is given by
\begin{equation}
\rho_e=\frac{\mu_e^3}{3\pi^2}.
\end{equation}
Similarly, the muon density can be written as
\begin{equation}
\rho_\mu=
\frac{\left(\mu_\mu^2-m_\mu^2\right)^{3/2}}
{3\pi^2},
\end{equation}
where the muon mass is taken as $m_\mu=105.7 \mathrm{MeV}$.

The total energy density of charge-neutral quark matter is obtained by adding the leptonic contributions to the energy density of the mPNJL model,
\begin{equation}
\mathcal{E} = \mathcal{E}_{\mathrm{mPNJL}} +\frac{\mu_e^4}{4\pi^2} + \frac{1}{\pi^2}\int_0^{(\mu_\mu^2-m_\mu^2)^{1/2}}
\hspace{-1cm}dk\,k^2(k^2+m_\mu^2)^{1/2},
\end{equation}
while the corresponding total pressure is
\begin{equation}
P = P_{\mathrm{mPNJL}} + \frac{\mu_e^4}{12\pi^2}
+ \frac{1}{3\pi^2}\int_0^{(\mu_\mu^2-m_\mu^2)^{1/2}}
\hspace{-0.8cm}\frac{dk\,k^4}{(k^2+m_\mu^2)^{1/2}}.
\end{equation}
These expressions incorporate the contributions of both electrons and muons, ensuring a self-consistent description of electrically neutral matter in beta equilibrium.

\section{Results and Discussions}
\label{Res}

We construct hybrid equations of state for cold hybrid stars by combining the SFHo, DD2, \mbox{DD2$_{hyp}$}, \mbox{NL3$\omega\rho$}~\cite{SteinerHempelFischer2013,Typel2010,Fortin:2017dsj,Horowitz:2000xj} models for the low-density hadronic phase with the mPNJL model for the high-density quark phase. The phase transition is implemented via the Maxwell construction~\cite{Alford:2004pf}. The deconfinement transition is also determined via the Maxwell construction between the $\Phi = 0$ and $\Phi = \Phi^{+}$ branches (see Eqs. (\ref{Eq:PL_sol1}) and (\ref{Eq:PL_sol3})). These equations of state are then used to determine the global properties of hybrid stars by solving the Tolman–Oppenheimer–Volkoff equations~\cite{Oppenheimer:1939ne,Tolman:1939jz}.

\subsection{The parameters of the mPNJL model\label{sfho}}

In the following, we discuss the effect of the different parameters of the modified Polyakov loop on the deconfinement phase transition and the properties of hybrid stars. This study will be performed implementing the SFHo model for the low density part of the hybrid equation of state~(EOS). We want to generate a stiff enough EOS that describes  two solar mass stars and, because of that, we introduced vector terms, in particular, the 4-quark  and the 8-quark vector~\cite{Ferreira:2020kvu} terms in the Lagrangian density. As will be shown below, we will also have to consider stiffer hadron EOS. A bag model parameter will be introduced when necessary to control the hadron-quark phase transition. For these particular cases, the equations of state of the mPNJL model will be replaced as follows
\begin{align}
\mathcal{E}_{\mathrm{mPNJL}} &\rightarrow \mathcal{E}_{\mathrm{mPNJL}} - B_0,
\end{align}
and
\begin{align}
P_{\mathrm{mPNJL}} &\rightarrow P_{\mathrm{mPNJL}} + B_0.
\end{align}

\noindent
The different parameters of the model will be varied within a range of values that allows for hybrid stars, i.e. neutron stars with a quark core,
and simultaneously predict maximum masses above two solar masses.

\subsubsection{Effect of $T_0$\label{sec:t0}}
\begin{figure}[!htb]
    \centering
    \includegraphics[width=\columnwidth]{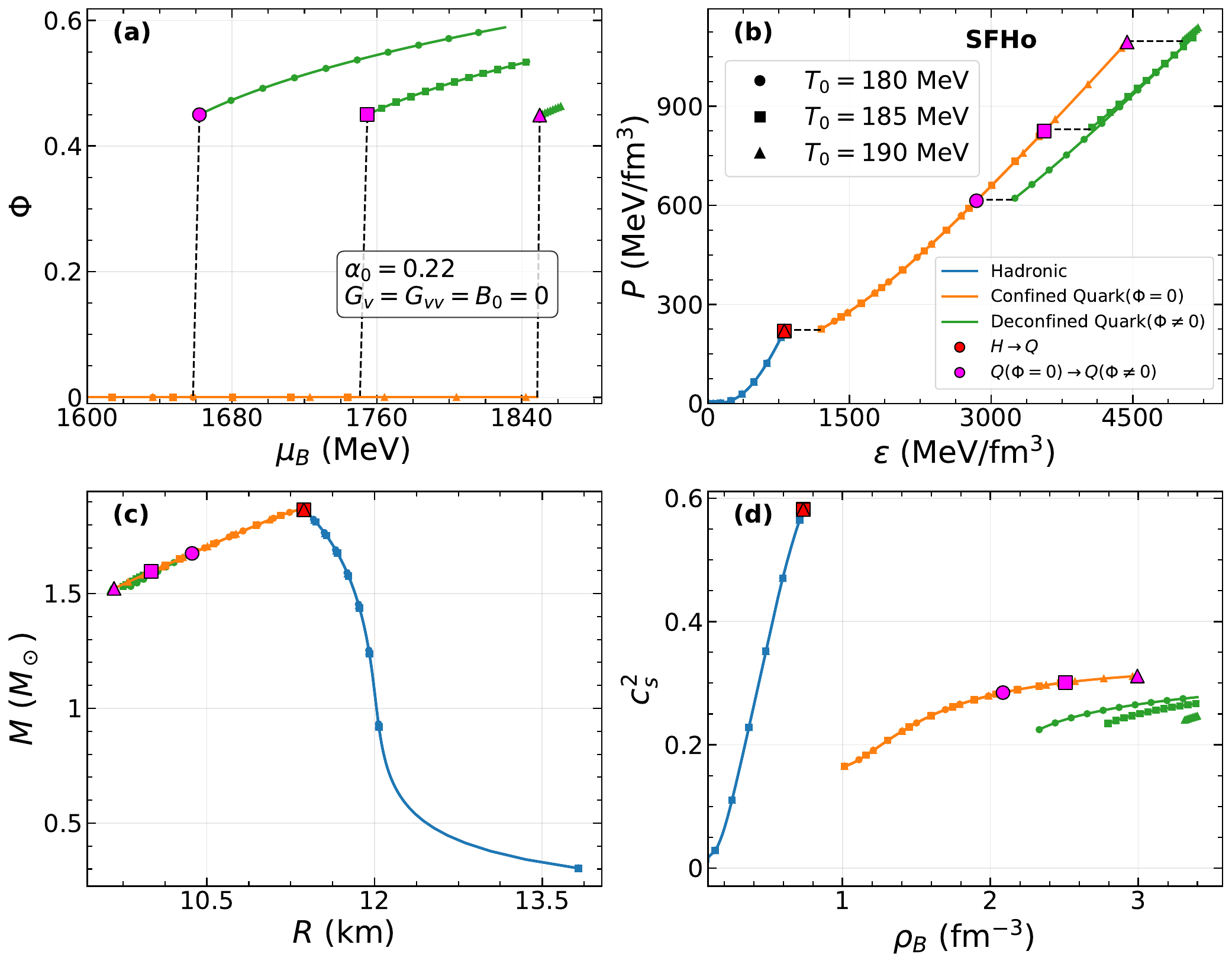}
    \caption{(a) Polyakov loop as a function of the baryonic chemical potential, (b) pressure versus energy density, (c) mass-radius diagrams of hybrid stars, and (d) squared speed of sound as a function of the baryonic density for various values of $T_0$.}
    \label{fig:EOS_MR_T0}
\end{figure}

First, we examine the effect of varying $T_0$ while fixing $\alpha_0=0.22$ and $G_V=G_{\v\v}=B_0=0$. Increasing $T_0$ shifts the confinement-deconfinement (\mbox{Q-Q}) transition to higher baryonic chemical potentials, $\mu_B$, as shown in Fig.~\ref{fig:EOS_MR_T0}{\color{blue}a}. 
 We also remark that, in the mPNJL model, the Polyakov loop presents nonzero solutions even in its simpler form, with $G_V=G_{\v\v}=0$, i.e., without the inclusion of vector channels. In another version of the PNJL model incorporating deconfinement phenomenology at $T=0$, presented in Refs.~\cite{Mattos_2019,Mattos:2021alf,Mattos2_2021}, the vector channel plays an important role in producing solutions with $\Phi>0$. The physical reason is that the Polyakov potential in that particular model, referred to as the PNJL0 model, also depends on the vector channel through the coupling constant $G_V$.

From Fig.~\ref{fig:EOS_MR_T0}{\color{blue}b}, we see that the increase of $T_0$ shifts the \mbox{Q-Q} transition to higher values of pressure. For this combination of fixed parameters the minimum value of $T_0 = 180$~MeV is allowed to achieve the hadron-quark (\mbox{H-Q}) phase transition. In Fig.~\ref{fig:EOS_MR_T0}{\color{blue}c}, the mass-radius (M-R) diagrams are presented. For all values of $T_0$ considered together with the fixed parameters, the \mbox{M-R}  branches become unstable as soon as \mbox{H-Q} transition takes place. No quark core stable hybrid star was found for these combinations. However, we emphasize that a more detailed radial oscillations analysis is not performed in this study. In Fig.~\ref{fig:EOS_MR_T0}{\color{blue}d} the squared speed of sound variation with baryonic density is shown in the three phases, namely,  hadronic, confined, and deconfined quark ones.

\subsubsection{Effect of $\alpha_0$}

\begin{figure}[!htb]
    \centering
    \includegraphics[width=\columnwidth]{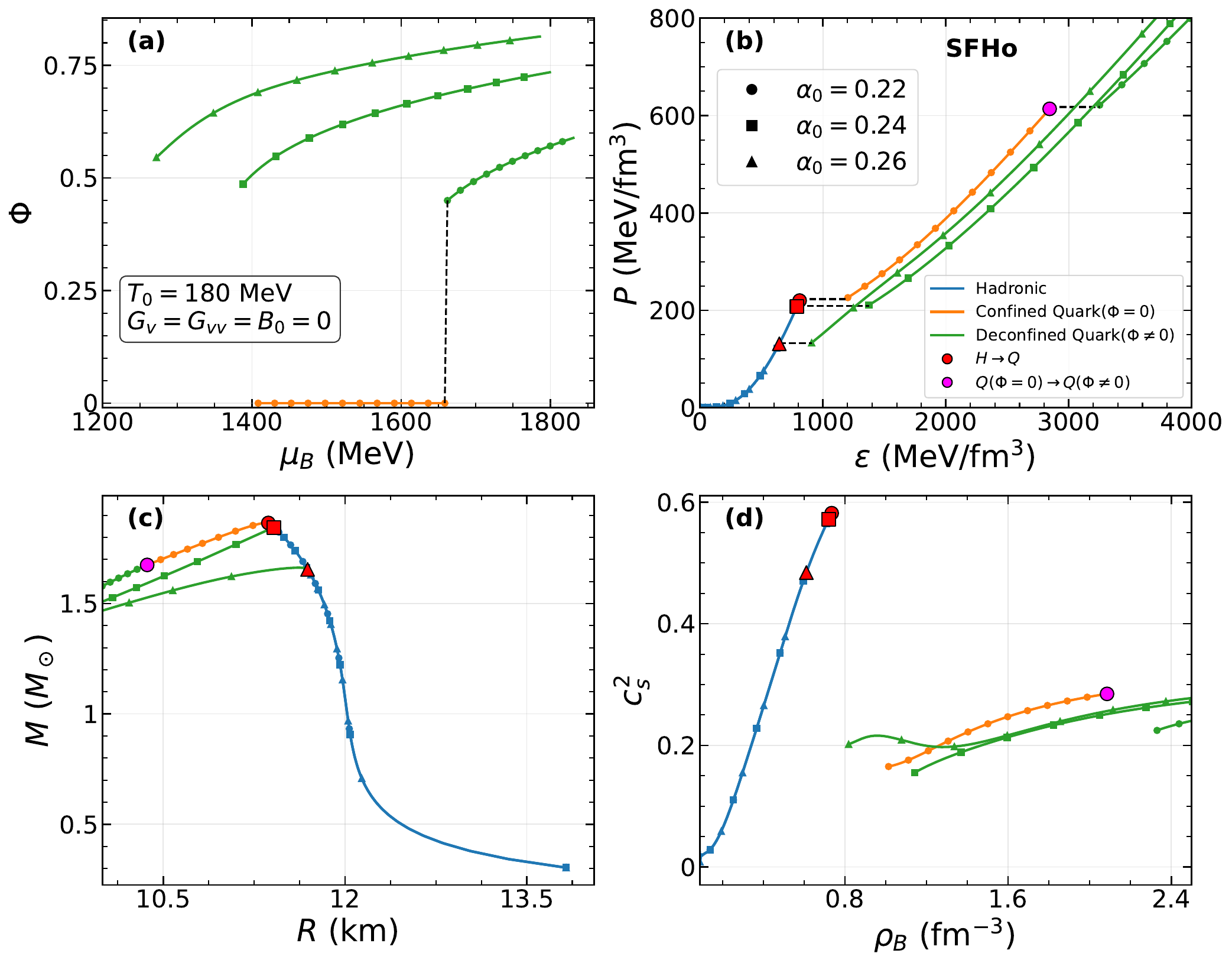}
    \caption{Same as Fig.~\ref{fig:EOS_MR_T0} but for different values of $\alpha_0$ with $T_0$ fixed.}
    \label{fig:EOS_MR_alpha0}
\end{figure}

Next, we investigate the effect of varying $\alpha_0$ while fixing $T_0=180$~MeV, $G_V = G_{\v\v} = B_0 = 0$. The value of $T_0$ was chosen because this value has shifted the \mbox{Q-Q}  transition to lower $\mu_B$ in the previous section. In contrast to increasing $T_0$, larger values of $\alpha_0$ shift the \mbox{Q-Q} transition to lower baryonic chemical potentials, $\mu_B$, since the parameter $\eta_2$ that determines the Polyakov loop strength decreases; see Fig.~\ref{fig:EOS_MR_alpha0}{\color{blue}a}. 
Consequently, the confined phase in the hybrid EOS decreases, as shown in Fig.~\ref{fig:EOS_MR_alpha0}{\color{blue}b}, and eventually disappears because the transition \mbox{H-Q} occurs directly to the deconfined quark phase: for $\alpha_0 = 0.24$, and $0.26$ the confined phase occurs before the \mbox{H-Q}  transition. As found above,  as soon as the  \mbox{H-Q} transition takes place, the M-R branch becomes unstable, i.e., no stable hybrid star with a quark core has been found, according to the findings presented in Fig.~\ref{fig:EOS_MR_alpha0}{\color{blue}c}.

\subsubsection{Effect of $\eta_2$}

\begin{figure}[!htb]
    \centering
    \includegraphics[width=\columnwidth]{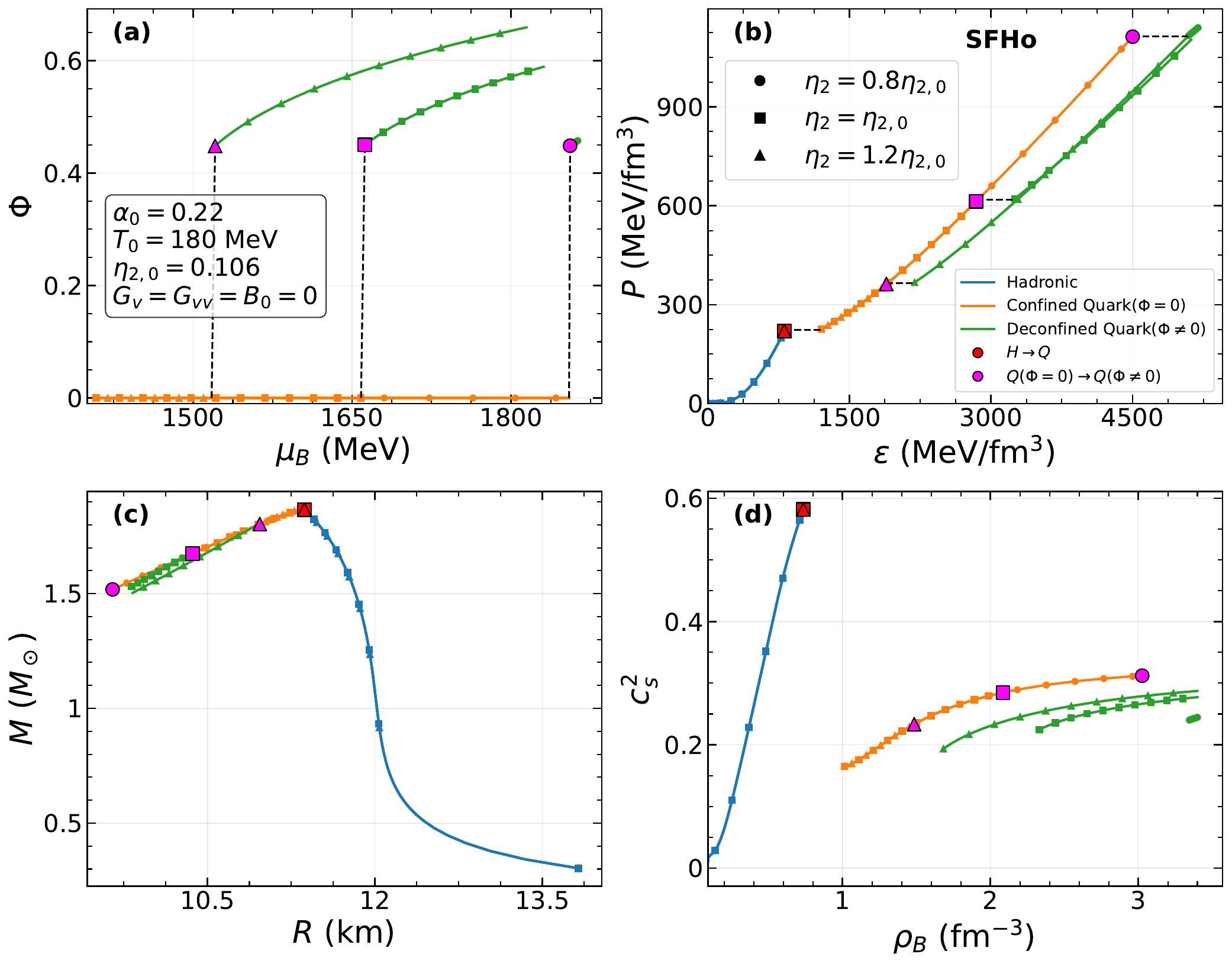}
    \caption{Same as Fig.~\ref{fig:EOS_MR_T0} but for different values of $\eta_2$.}
    \label{fig:EOS_MR_eta2}
\end{figure}

In the previous subsections, the parameter $\eta_2$ of the Polyakov loop has been defined by Eq.~\eqref{Eq:eta2}. In the following, we consider this quantity as an effective parameter and discuss its role when it is increased/decreased above/below the value defined by Eq.~\eqref{Eq:eta2}. Starting from the value of \(\eta_2\) defined by  setting \(T_0=180\) MeV, \(\alpha_0=0.22\), $G_V = G_{\v\v} = B_0 =0$, denoted here by $\eta_{2,0}$, we relax the relation between $\eta_2$ and these parameters. From this procedure, we verify that the main effect caused by changing $\eta_2$ is the shift of the \mbox{Q-Q} transition to larger~(smaller) densities when $\eta_2$ is reduced~(increased), see Fig.~(\ref{fig:EOS_MR_eta2}), since this corresponds to reducing (increasing) the strength of the Polyakov loop potential. As discussed in the previous cases,  the M-R branches become unstable at the \mbox{H-Q} transition, and no  stable hybrid star was found with a quark core.

\subsubsection{Effect of $G_V$}

\begin{figure}[!htb]
    \centering
    \includegraphics[width=\columnwidth]{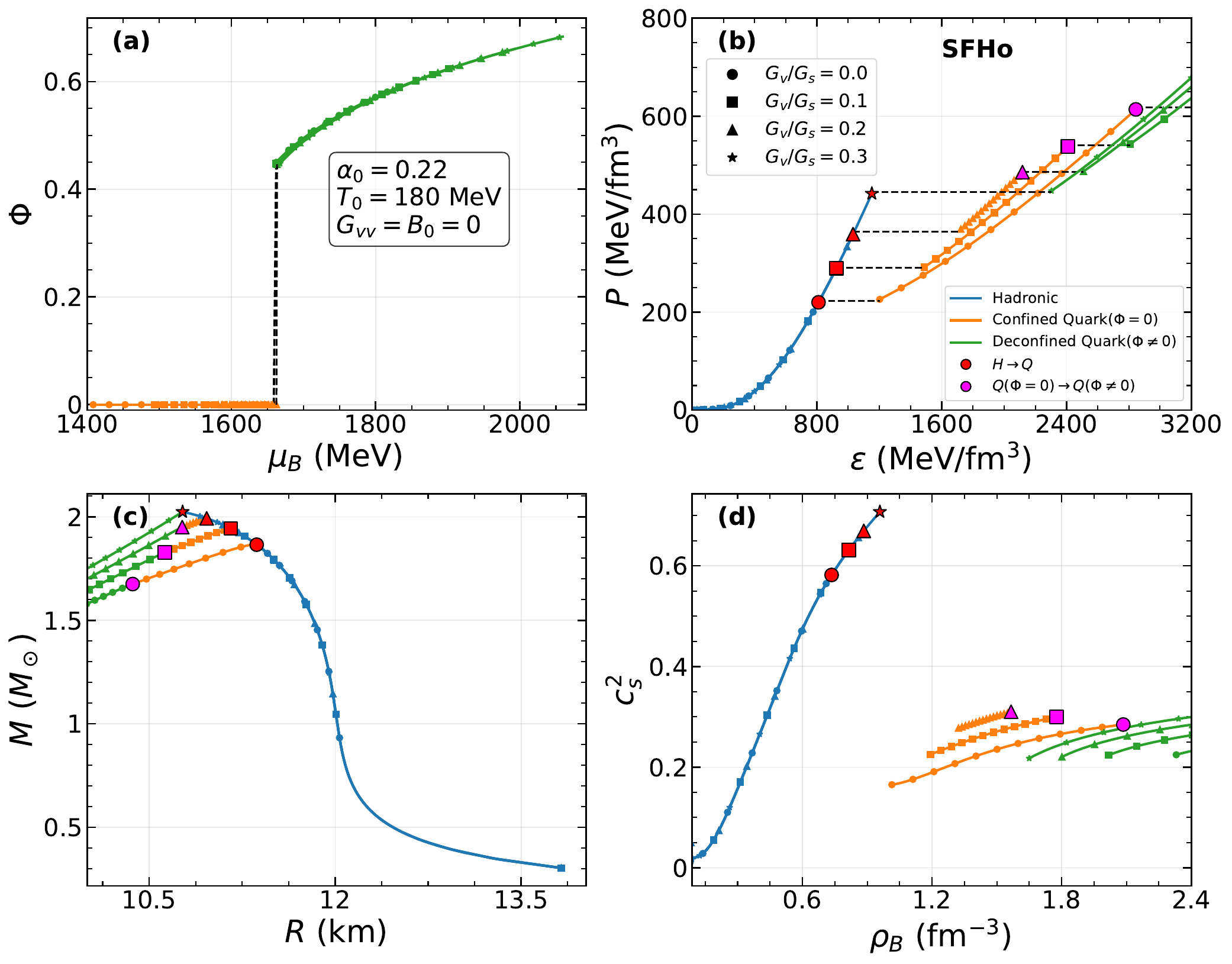}
    \caption{Same as Fig.~\ref{fig:EOS_MR_T0} but for different values of $G_V$.}
    \label{fig:EOS_MR_GV}
\end{figure}

To obtain stable hybrid star branches with a quark core, we next introduce repulsive vector interactions governed by the coupling constant $G_V$, while initially fixing $T_0=180$~MeV, $\alpha_0=0.22$, and $G_{\v\v} = B_0 =0$, see Fig.~\ref{fig:EOS_MR_GV}. 
 Fig.~\ref{fig:EOS_MR_GV}{\color{blue}a} shows that the effect of $G_V$ on the Polyakov loop is only marginal, in contrast to the PNJL0 model of Refs.~\cite{Mattos_2019,Mattos:2021alf,Mattos2_2021}, where increasing $G_V$ shifts the transition chemical potential toward lower values. However, the increase in $G_V$ shifts the \mbox{H-Q} transition to higher pressure and energy density and shifts the \mbox{Q-Q} transition to lower pressure and energy density. Hence, it reduces the confined phase and at $G_V = 0.3G_s$, the transition becomes \mbox{H-Q}~($\Phi\neq0$), and no confined phase occurs  as we show in Fig.~\ref{fig:EOS_MR_GV}{\color{blue}b}. 
From Fig.~\ref{fig:EOS_MR_GV}{\color{blue}c}, it is evident that as soon as the \mbox{H-Q} transition occurs, the M-R branch becomes unstable. As a remark, notice that now it is possible to find at least one configuration in which the star mass is equal to $2M_\odot$, differing from the previous diagrams in which $G_V=0$.

\subsubsection{Effect of $G_{\v\v}$}

\begin{figure}[!htb]
    \centering
    \includegraphics[width=\columnwidth]{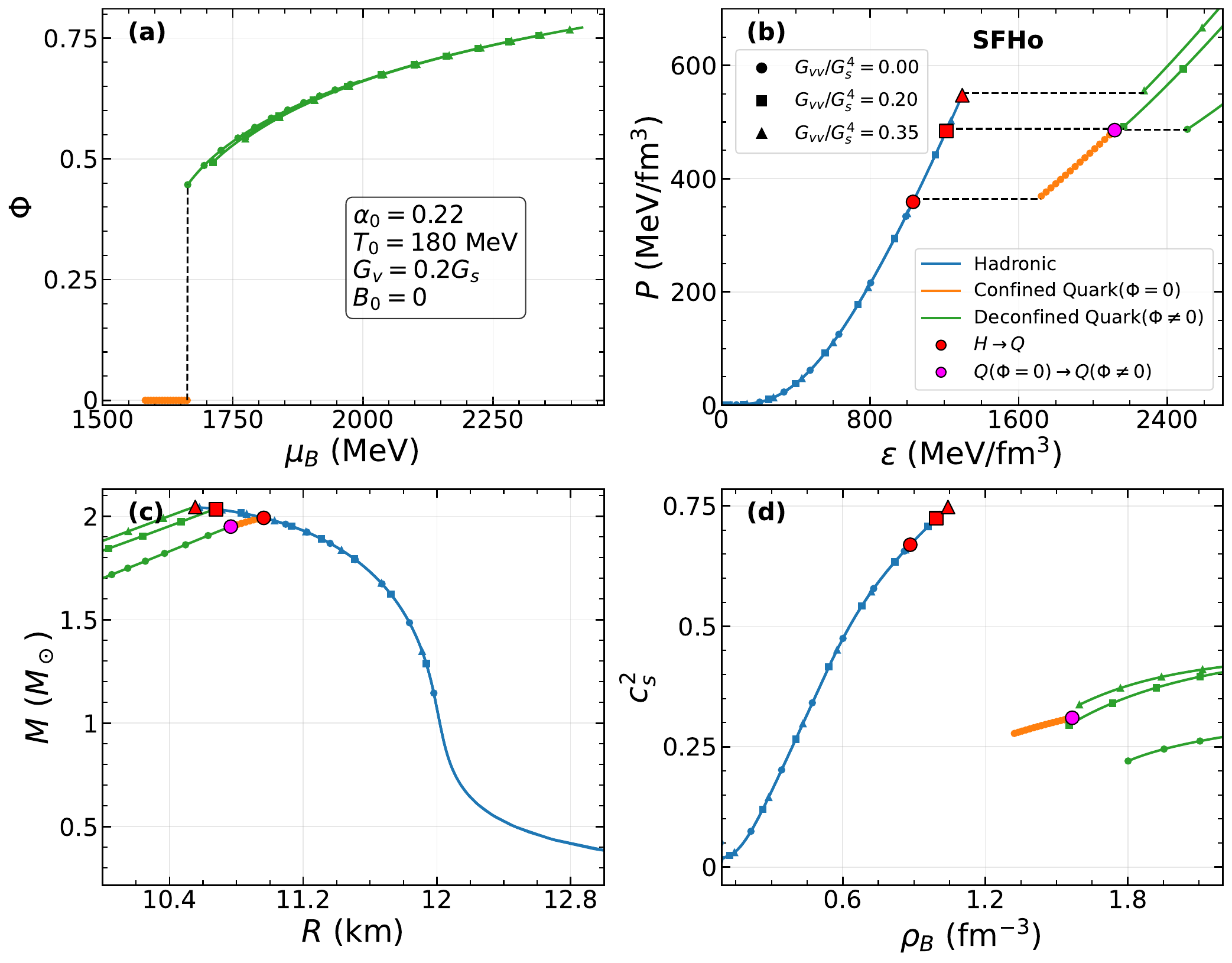}
    \caption{Same as Fig.~\ref{fig:EOS_MR_T0} but for different values of $G_{\v\v}$.}
    \label{fig:EOS_MR_Gvv}
\end{figure}

We now investigate how the phase transition is affected by the \mbox{8-quark} repulsive interaction coupling $G_{\v\v}$, which helps achieve a quark core hybrid star. The results are displayed in Fig.~\ref{fig:EOS_MR_Gvv}. 
Increasing  $G_{\v\v}$ shifts the \mbox{H-Q} transition to higher pressure and energy density and shifts the \mbox{Q-Q} transition to lower density, and eventually the $\Phi=0$ phase disappears above $G_{\v\v}/G_s^4=0.20$. Similar to the previous cases, as soon as the \mbox{H-Q} transition takes place, the M-R branches become unstable.

\subsubsection{Effect of $B_0$}

\begin{figure}[!htb]
    \centering
    \includegraphics[width=\columnwidth]{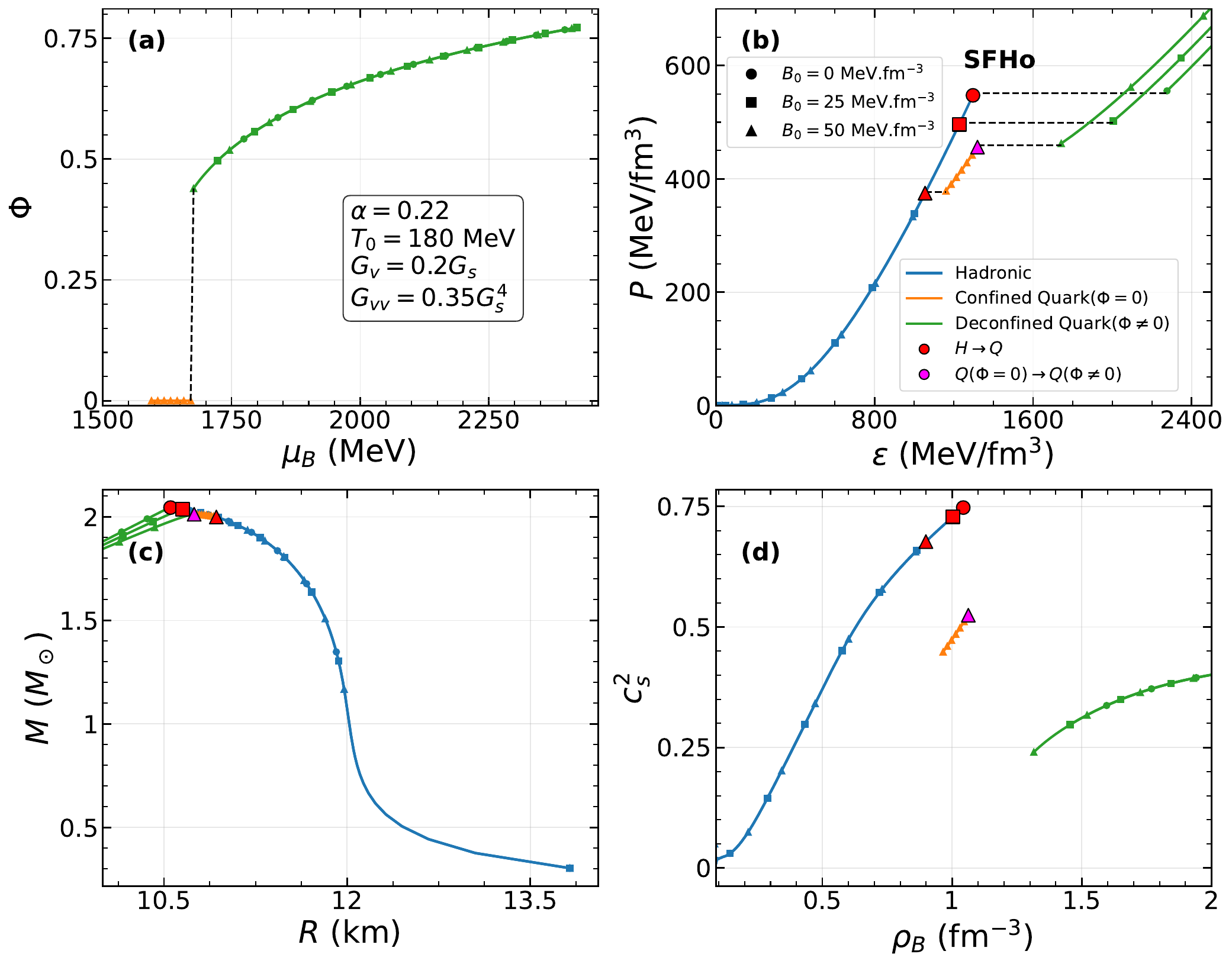}
    \caption{Same as Fig.~\ref{fig:EOS_MR_T0} but for different values of $B_0$.}
    \label{fig:EOS_MR_B0}
\end{figure}

Next, we study the effect of bag constant $B_0$ which moves the \mbox{H-Q} transition to lower $\mu_B$. We introduce the bag pressure $B_0$ on the   $G_{\v\v}/G_s^4=0.35$ case of Fig.~\ref{fig:EOS_MR_Gvv}. From Figs.~\ref{fig:EOS_MR_B0}{\color{blue}a} and~\ref{fig:EOS_MR_B0}{\color{blue}b}, we conclude that the increase of $B_0$ shifts the \mbox{H-Q} transition to lower energy density, and at $B_0 = 50$~\mbox{MeV fm$^{-3}$} the confined phase appears. From Fig.~\ref{fig:EOS_MR_B0}{\color{blue}c} it is seen that with $B_0 = 50$~\mbox{MeV fm$^{-3}$} a hybrid star with a  small confined core is obtained with a maximum mass $\sim 2 M_\odot$. At the center of the maximum mass hybrid star, the squared speed of sound exceeds the conformal limit, see Fig.~\ref{fig:EOS_MR_B0}{\color{blue}d} which is expected at density far below the perturbative-QCD density.
From the above analysis, we expect to achieve  massive cold hybrid stars with a quark core by conveniently choosing the parameters of the model. We have described the hadron phase with the SFHo EOS, which is a soft EOS. In the following, we will consider stiffer EOS and will choose a parametrization of the mPNJL model so that stable hybrid stars with a quark core exceeding $\sim 2 M_\odot$ are obtained.

\subsection{Massive hybrid stars}

Having understood the effect of the different parameters of the mPNJL model, in the next subsections, we propose several hybrid star EOSs that contain either a deconfined quark phase core or a confined core in the center of the NS and have a mass $\gtrsim 2M_\odot$. This will be possible by choosing the parameters conveniently and describing the hadron phase with an EOS  that is stiffer than SFHo. We consider DD2~\cite{Typel2010}, \mbox{DD2$_{hyp}$}~\cite{Fortin:2017dsj} and \mbox{NL3$\omega\rho$}~\cite{Horowitz:2000xj}.

\subsubsection{With deconfined quark core}

\begin{figure}[!htb]
    \centering
    \includegraphics[width=\columnwidth]{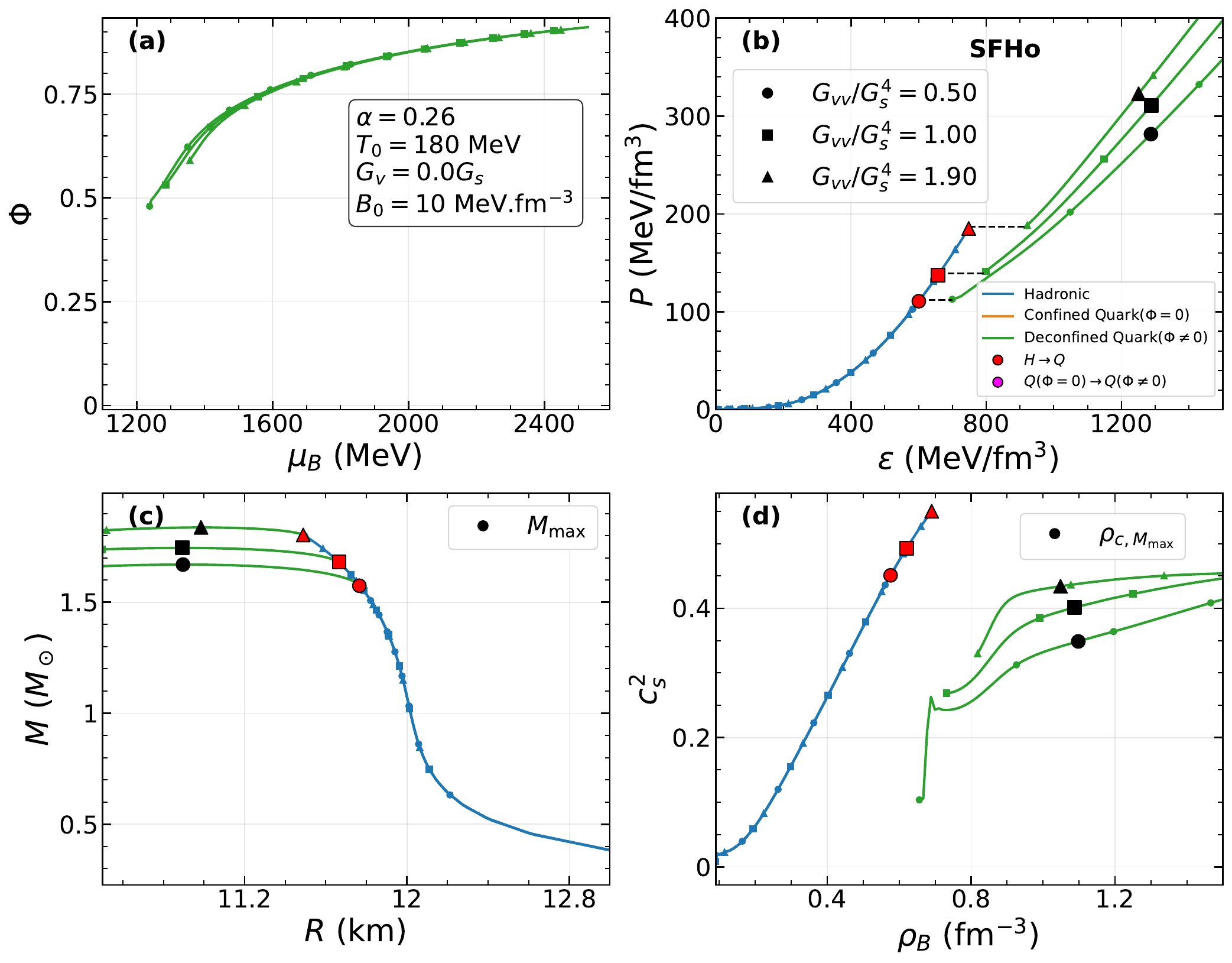}
    \caption{(a) Polyakov loop as a function of the baryonic chemical potential, (b) pressure versus energy density, (c) mass-radius diagrams of hybrid stars with deconfined quark cores, and (d) squared speed of sound as a function of the baryonic density. The SFHo model is used in the hadronic sector.}
    \label{fig:Deconf_sfho}
\end{figure}

\begin{figure}[!htb]
    \centering
    \includegraphics[width=\columnwidth]{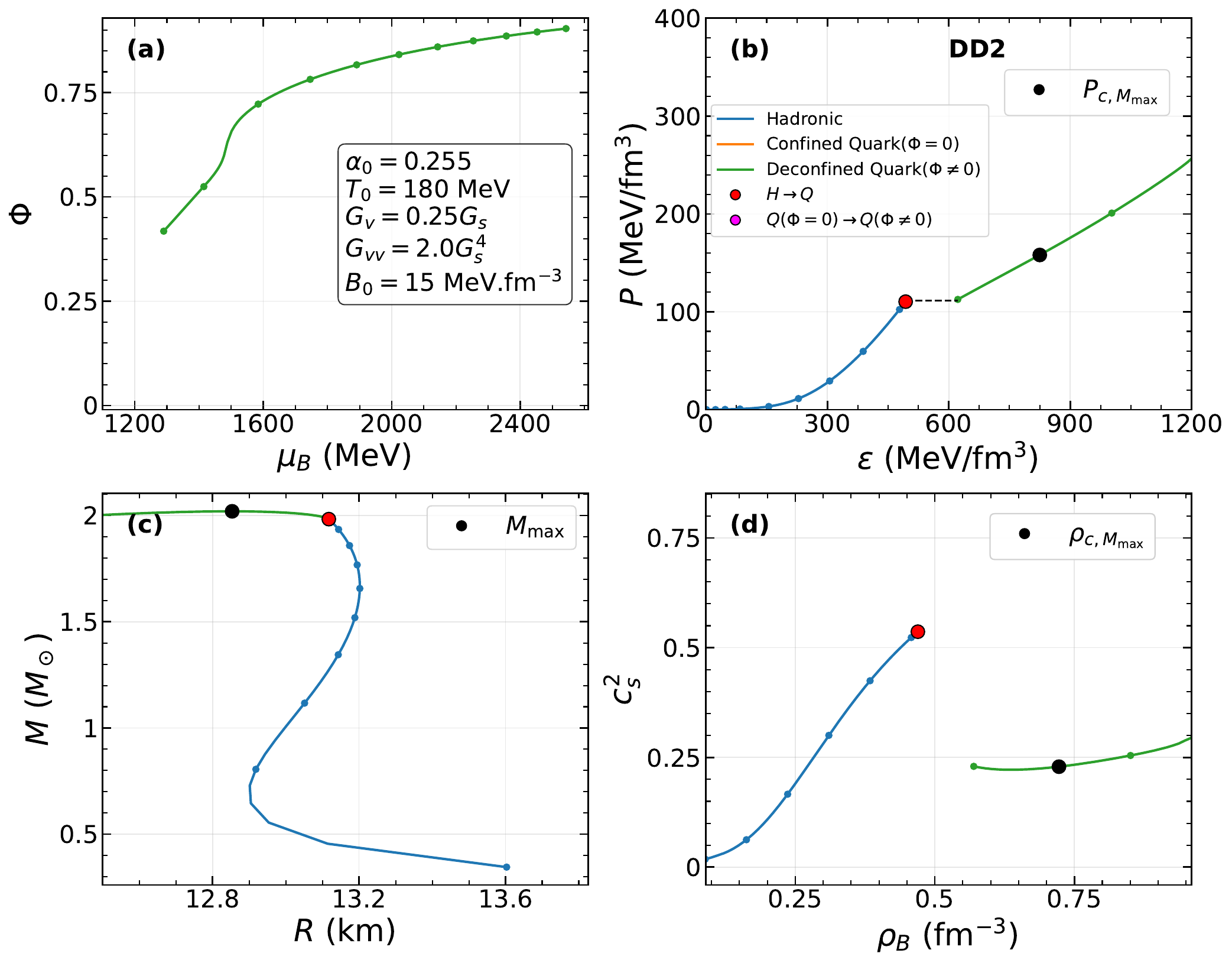}
    \caption{The same as in Fig.~\ref{fig:Deconf_sfho}, but for the DD2 model in the hadronic sector, and a different parameter set of the mPNJL model.}
    \label{fig:Deconf_DD2}
\end{figure}

From the previous analysis, it is found that increasing $\alpha_0$ or decreasing $T_0$ shifts the \mbox{Q-Q} transition to a lower chemical potential $\mu_B$, and increasing the coupling  $G_V$ or the  bag constant $B_0$ shifts the \mbox{H-Q} transition to a lower $\mu_B$. We choose $\alpha_0 =0.26$, $G_V =0$, $T_0 =180$~MeV, and $B_0=10~\mbox{MeV fm}^{-3}$, which shift the \mbox{Q-Q} transition below the  \mbox{H-Q}~($\Phi\neq0$), so there are only two phases: (i)~a hadron phase~(blue), and (ii)~a deconfined quark phase~(green), procedure also adopted in the hadron-quark phase transition studied with the PNJL0 model in Ref.~\cite{Mattos2_2021}. This combination of parameters allows us to achieve  massive hybrid stars that  contain deconfined quark matter in their cores. We can see from Fig.~\ref{fig:Deconf_sfho} that the phase transition occurs directly from hadron matter to a  deconfined quark phase, and it happens well inside the stable hybrid stars. 
However, having considered the  SFHo model for the hadron phase, the maximum masses of the stable branches are below $2M_\odot$. To have deconfined quark core hybrid stars with masses on the order of two solar masses,  we replace the SFHo EOS with the DD2 EOS, a stiffer model at low density. The results are shown in Fig.~\ref{fig:Deconf_DD2}. 
With this choice of low density EOS and  a specific parameter set of the mPNJL model, the \mbox{H-Q} transition is taking place at a lower $\mu_B$, see Fig.~\ref{fig:Deconf_DD2}{\color{blue}a}, and the maximum mass is above $2M_\odot$ with a reasonable amount of deconfined quark core, as displayed in Fig.~\ref{fig:Deconf_DD2}{\color{blue}c}.

\begin{figure}[!htb]
    \centering
    \includegraphics[width=0.75\linewidth]{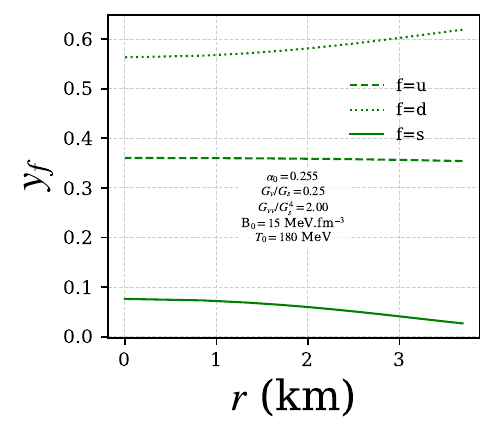}
    \caption{Variations of individual quark fractions, $y_f=\rho_f/\rho$, inside the maximum mass hybrid star in the stable branch as a function of radial distance $r$ from the center towards the surface up to the hadron-quark phase transition~($\Phi\ne0$) of Fig.~\ref{fig:Deconf_DD2}{\color{blue}c}.}
    \label{fig:Deconf_DD2_frac}
\end{figure}

\begin{figure}[!htb]
    \centering
    \includegraphics[width=0.8\columnwidth]{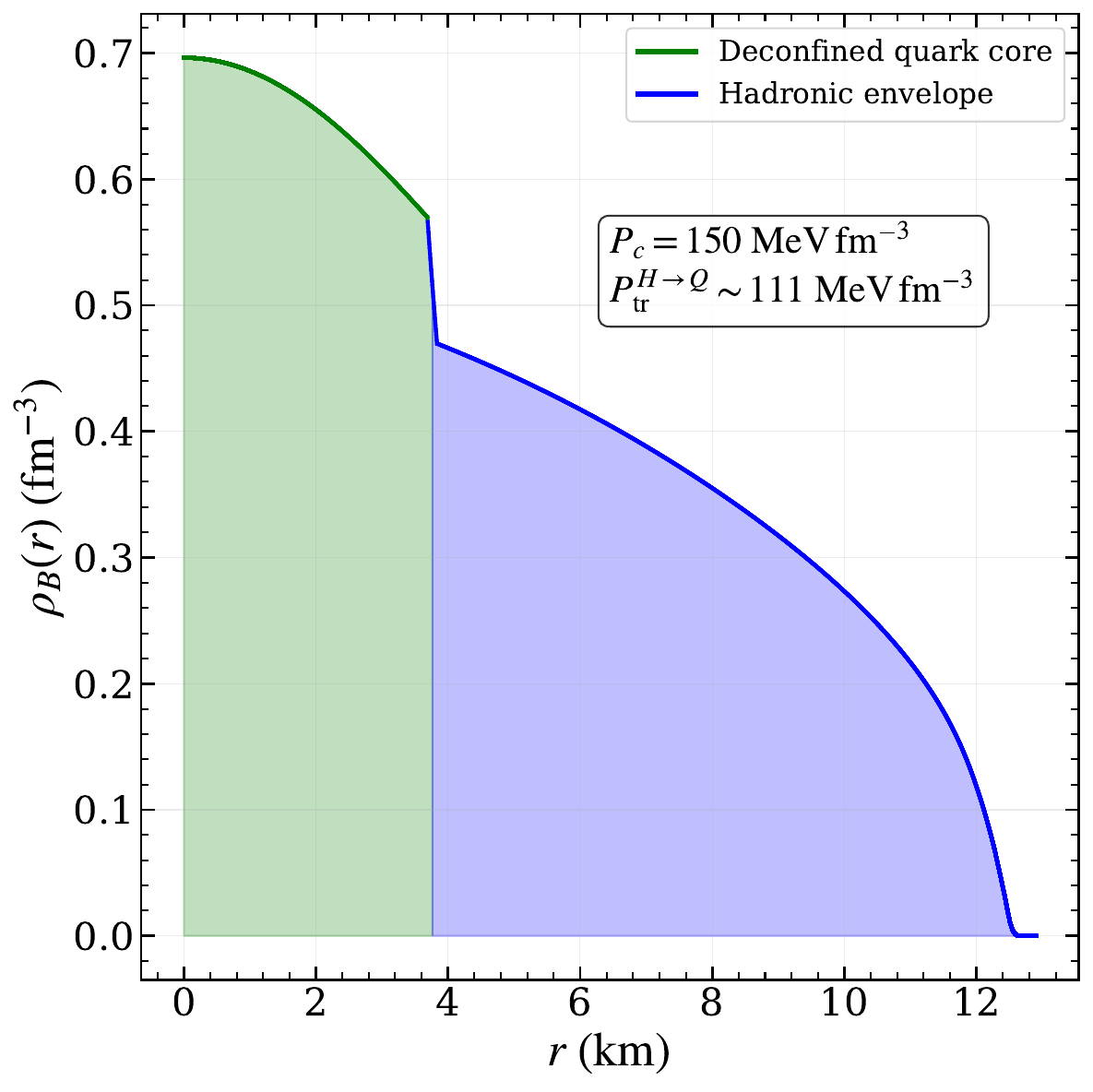}
    \caption{Baryonic density profile as a function of radial distance for the stable maximum mass hybrid star  of Fig.~\ref{fig:Deconf_DD2}{\color{blue}c}.}
    \label{fig:density_prof_deconf_DD2}
\end{figure}

 Now we present some results regarding the star with a maximum mass of Fig.~\ref{fig:Deconf_DD2}{\color{blue}c}, represented by the black circle. In Fig.~\ref{fig:Deconf_DD2_frac} the variation of quark fraction inside this particular star from the center towards the surface is displayed. 
The strange quark is always present, but still in small amounts due to its larger constituent mass. In Fig.~\ref{fig:density_prof_deconf_DD2}, we represent the density profile of this star, identifying the  quark and hadron phases in $\beta$-equilibrium with electrons and muons. 
The deconfined quark matter core  has a radius of $\sim$ 4 km. Note that the blue region consists of nucleons with leptons and also includes the crust of the star.

\subsubsection{With confined quark core}

\begin{figure}[!htb]
    \centering
    \includegraphics[width=\columnwidth]{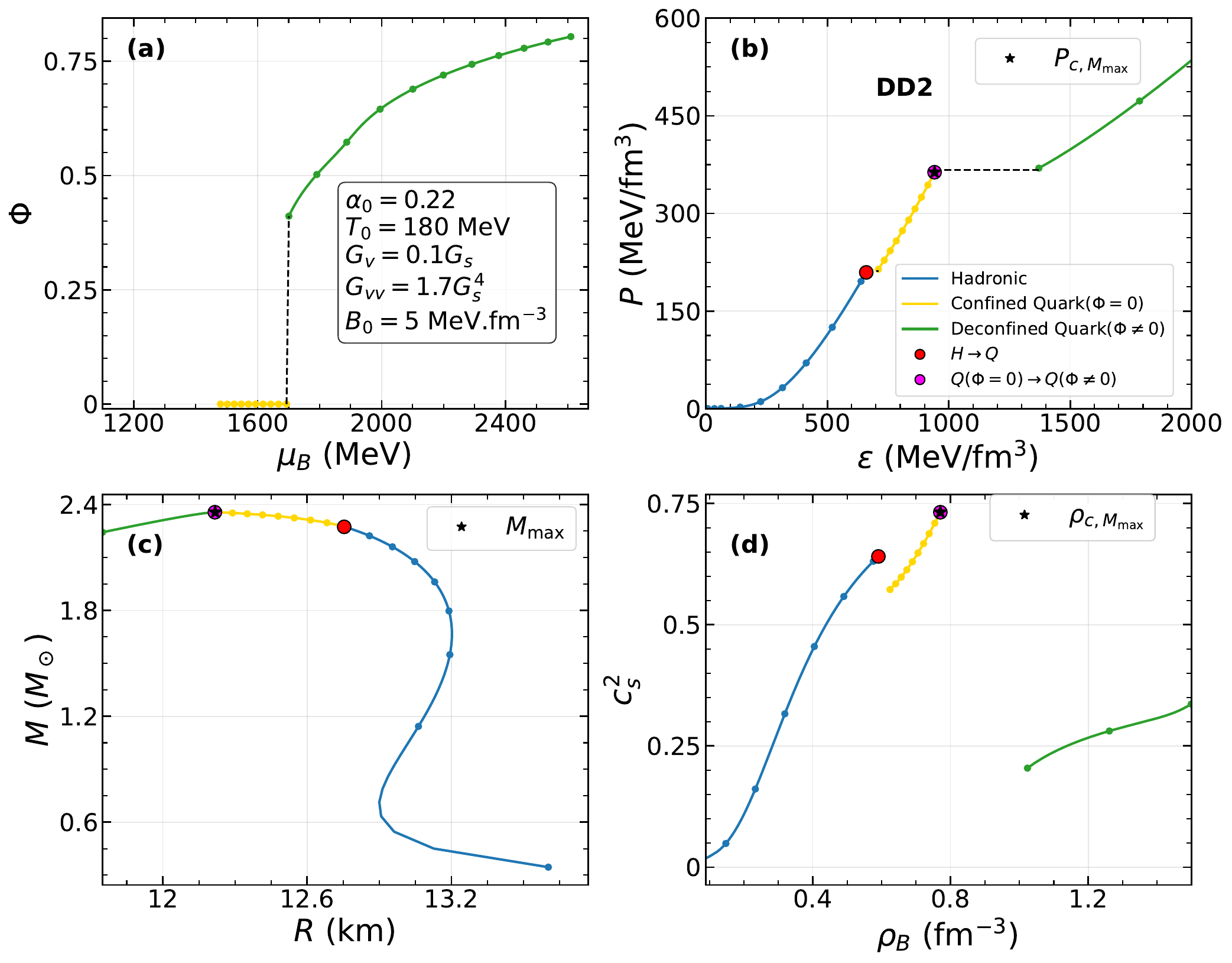}
    \caption{(a)~Polyakov loop as a function of the baryonic chemical potential, (b)~pressure versus energy density, (c)~mass-radius diagrams of hybrid stars with confined quark cores, and (d)~squared speed of sound as a function of the baryonic density. The DD2 model is used in the hadronic sector.}
    \label{fig:Quarkyonic_dd2}
\end{figure}

\begin{figure}[!htb]
    \centering
    \includegraphics[width=\columnwidth]{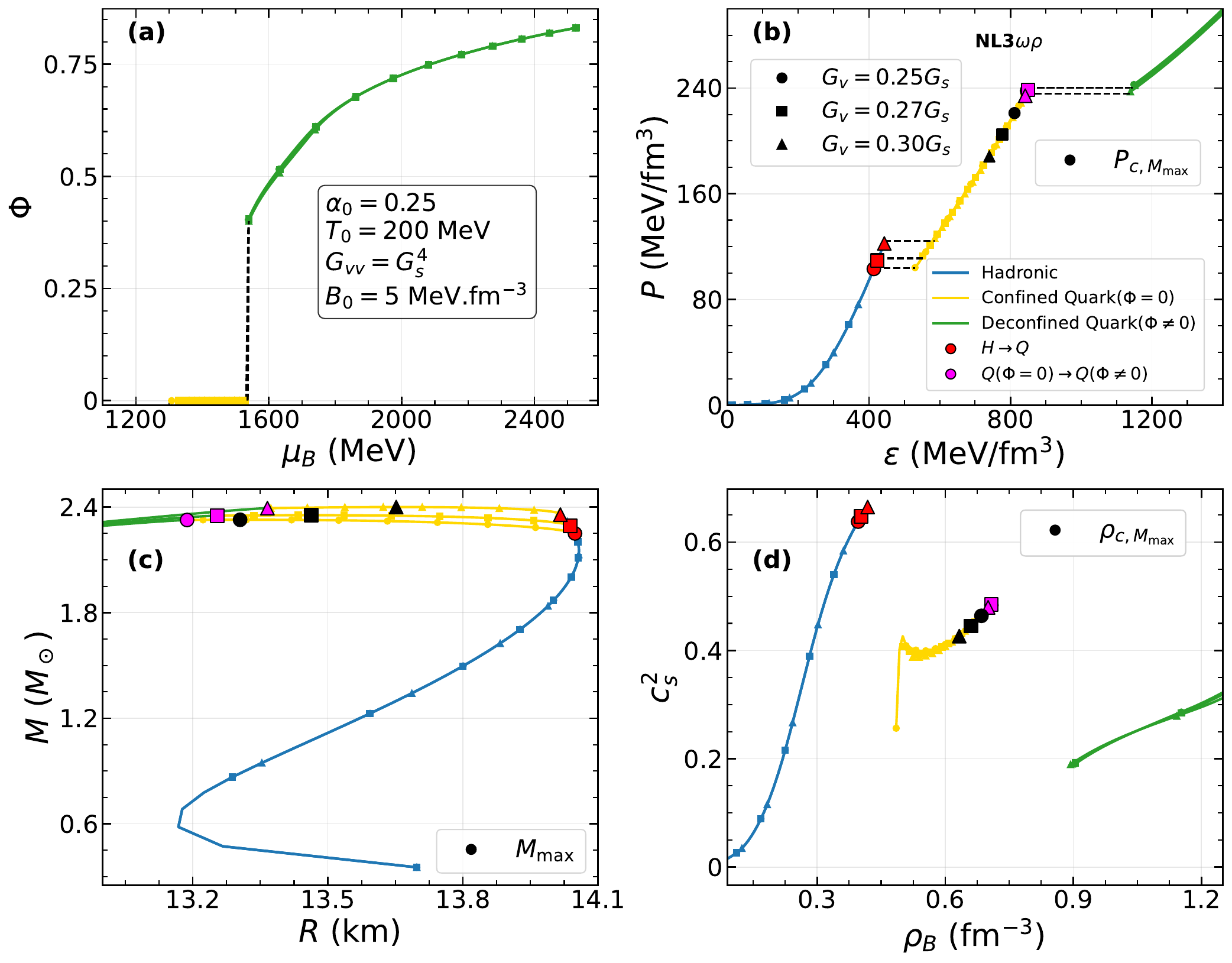}
    \caption{The same as in Fig.~\ref{fig:Quarkyonic_dd2}, but for the \mbox{NL3$\omega\rho$} model in the hadronic sector, and a different parameter set of the mPNJL model.}
    \label{fig:EOS_MR_nl3}
\end{figure}

\begin{figure}[!htb]
    \centering
    \includegraphics[width=\columnwidth]{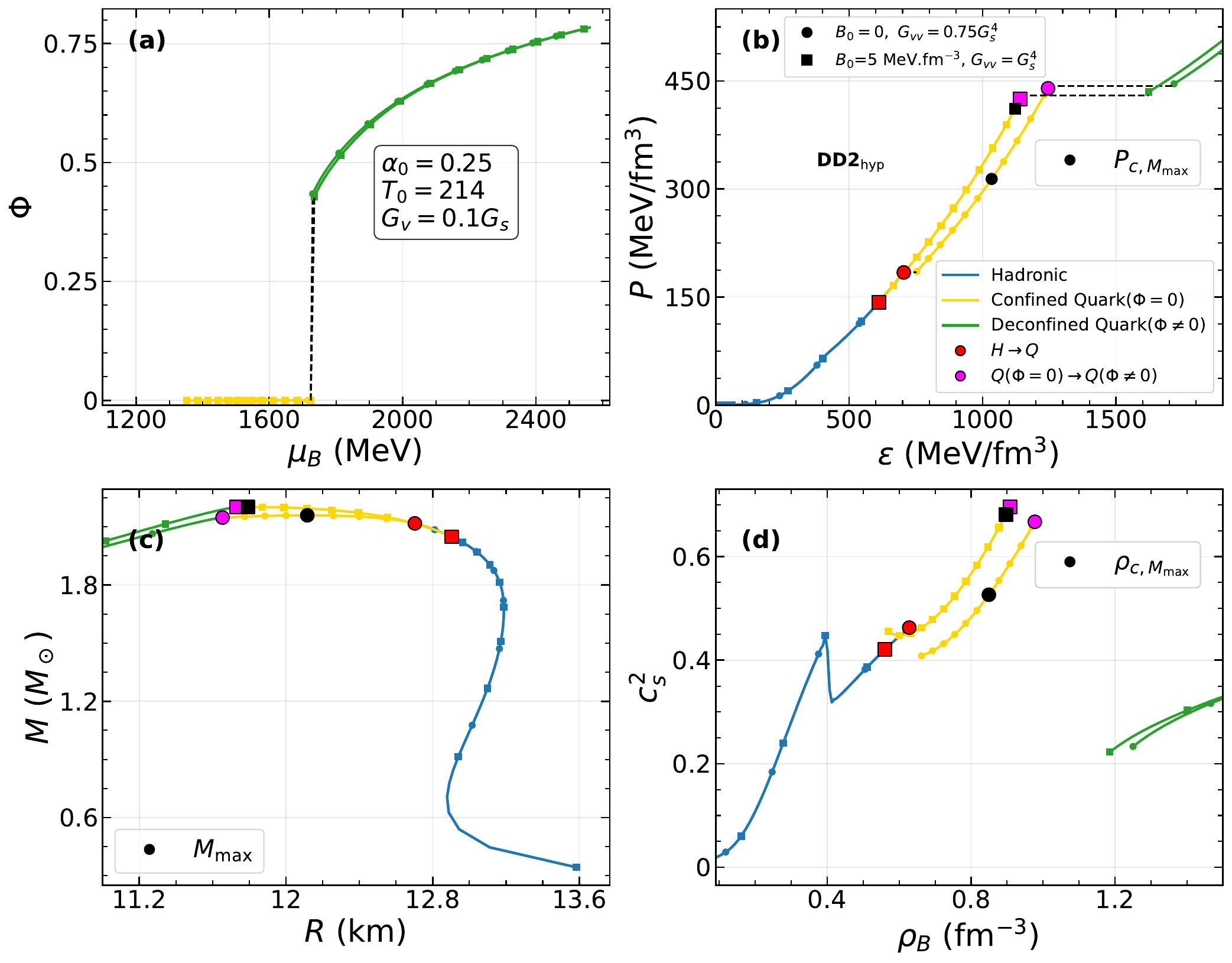}
    \caption{The same as in Fig.~\ref{fig:Quarkyonic_dd2}, but for the \mbox{DD2$_{hyp}$} model in the hadronic sector, and a different parameter set of the mPNJL model.}
    \label{fig:Quarkyonic_dd2Hyp}
\end{figure}

In this subsection, we discuss hybrid stars with a confined quark core. We first consider the DD2 model for the low density hadron phase. The results are shown in Fig.~\ref{fig:Quarkyonic_dd2}. Notice that the hadron to confined quark transition is taking place with an  extremely small  gap in the energy density, see Fig.~\ref{fig:Quarkyonic_dd2}{\color{blue}b}. 
Furthermore, the hybrid stars above $2M_\odot$ contain a confined quark core, as one can see in Fig.~\ref{fig:Quarkyonic_dd2}{\color{blue}c}. As soon as the deconfined quark phase sets in, the star becomes unstable. The speed of sound squared inside confined quark core stars is above the conformal limit to support massive stars, according to the findings of Fig.~\ref{fig:Quarkyonic_dd2}{\color{blue}d}. Similar results are found with two other low density EOSs as shown in Figs.~\ref{fig:EOS_MR_nl3} and~\ref{fig:Quarkyonic_dd2Hyp}.

\begin{figure}[!htb]
    \centering
    \includegraphics[width=0.8\columnwidth]{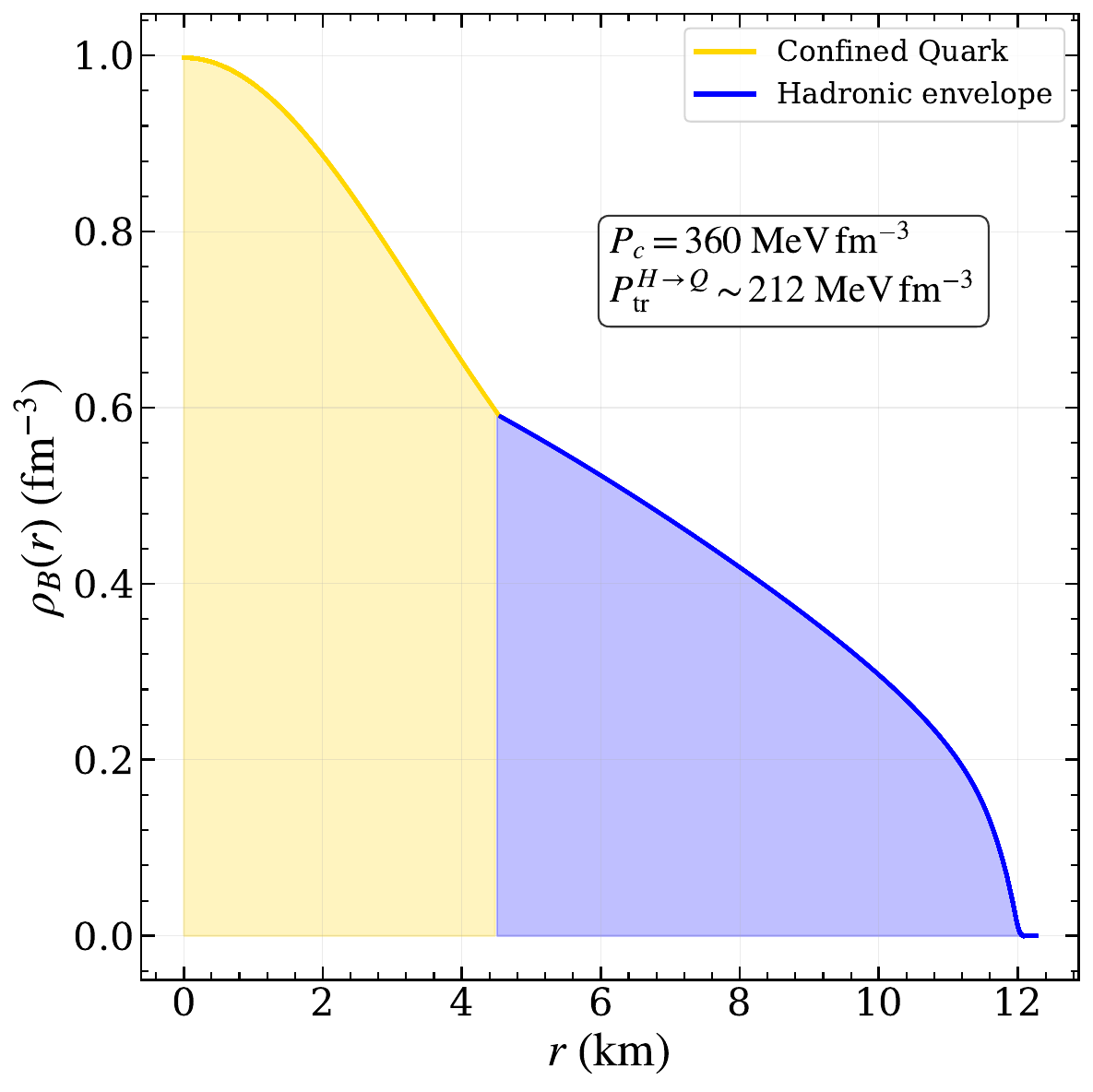}
    \caption{Baryonic density profile as a function of radial distance for the stable maximum mass hybrid star of Fig.~\ref{fig:Quarkyonic_dd2}{\color{blue}c}.}
    \label{fig:Quarkyonic_density_profile}
\end{figure}

In Fig.~\ref{fig:Quarkyonic_density_profile}, a similar density profile of the  confined quark core is shown as in Fig.~\ref{fig:density_prof_deconf_DD2}. 
It corresponds to a hybrid star with central pressure $P_{c,M_{max}}=360$~MeV fm$^{-3}$ for the parameter set of Fig.~\ref{fig:Quarkyonic_dd2}. The confined quark core has a radius of around~$5$~km.

\subsubsection{With both confined and deconfined quark cores}

\begin{figure}[!htb]
    \centering
    \includegraphics[width=\columnwidth]{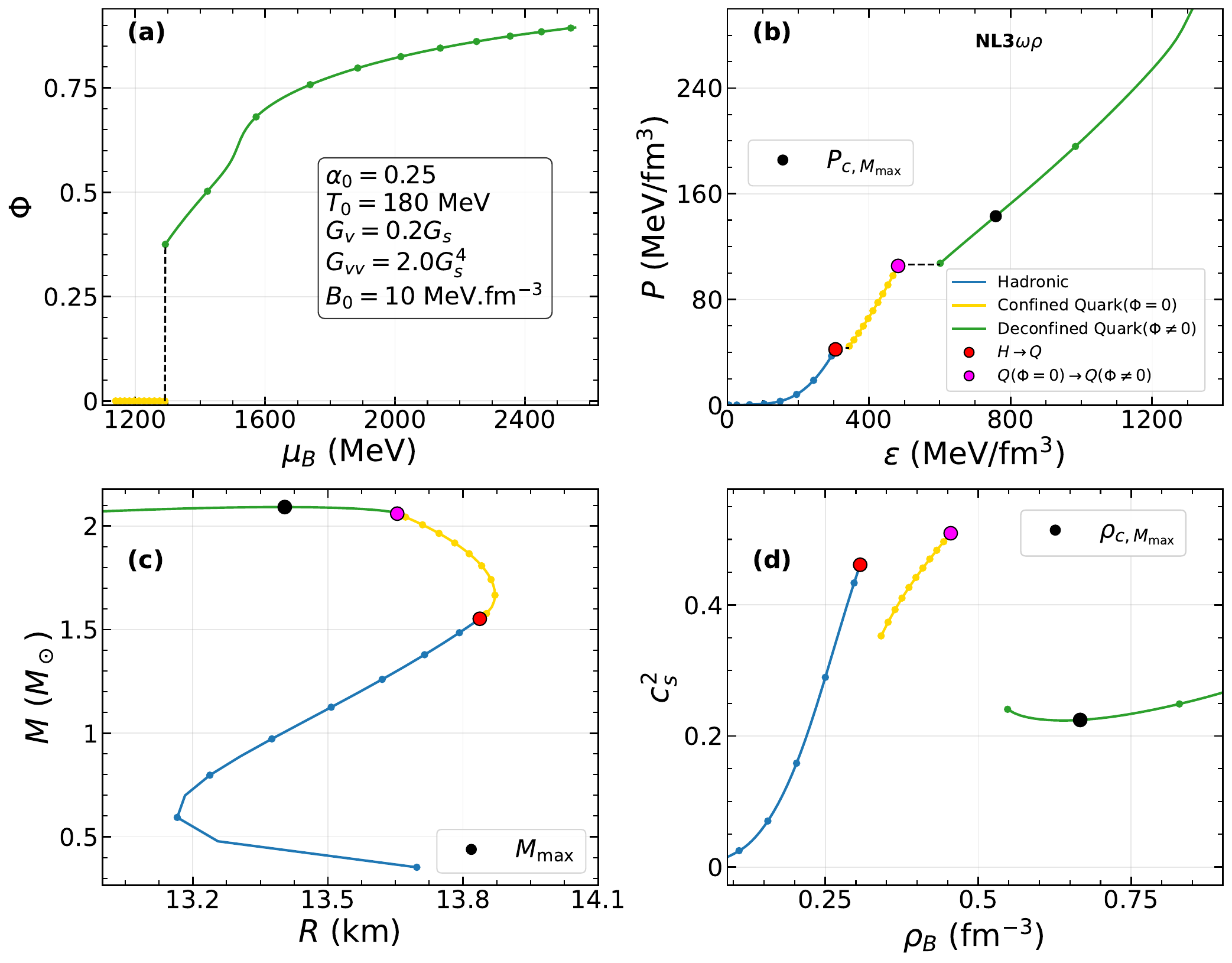}
    \caption{(a)~Polyakov loop as a function of the baryonic chemical potential, (b)~pressure versus energy density, (c)~mass-radius diagrams of hybrid stars with confined and deconfined quark cores, and (d)~squared speed of sound as a function of the baryonic density. The \mbox{NL3$\omega\rho$} model is used in the hadronic sector.}
    \label{fig:HQD_NL3WR}
\end{figure}

In this subsection, we will discuss the scenario of a stable hybrid star with a  mass above $2M_\odot$ that has  three phases:~(i)~hadronic, (ii)~confined quark, and (iii)~deconfined quark. It is achieved with the stiff EOS \mbox{NL3$\omega\rho$} and by tuning the mPNJL model parameters; see Fig.~\ref{fig:HQD_NL3WR}. 
From Fig.~\ref{fig:HQD_NL3WR}{\color{blue}b}, we see that the energy density gaps from one phase to the other are small. Also, we found that the speed of sound squared at the center of the  maximum mass star is below the conformal limit at the deconfined core, see Fig.~\ref{fig:HQD_NL3WR}{\color{blue}d}. This behavior 
is expected at finite baryon densities ($\rho_B \sim 2-5\rho_0$) where QCD is not conformal. The maximum mass for this three phase hybrid star is above $2.1M_\odot$ approximately.

\begin{figure}[!htb]
    \centering
    \includegraphics[width=0.75\linewidth]{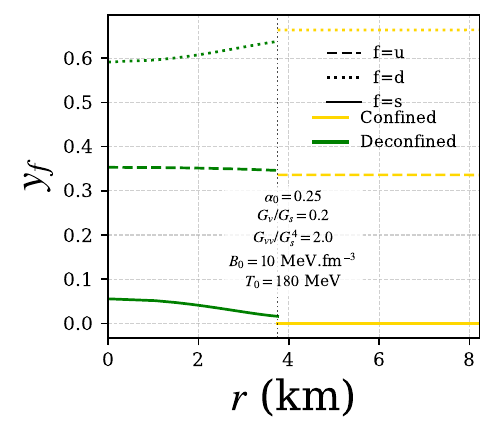}
    \caption{Variations of individual quark fractions, $y_f=\rho_f/\rho$, inside the maximum mass hybrid star in the stable branch as a function of radial distance $r$ from the center towards the surface up to the hadron-quark phase transition~($\Phi = 0$) of Fig.~\ref{fig:HQD_NL3WR}{\color{blue}c}.}
    \label{fig:HQD_NL3WR_frac}
\end{figure}

\begin{figure}[!htb]
    \centering
    \includegraphics[width=0.8\columnwidth]{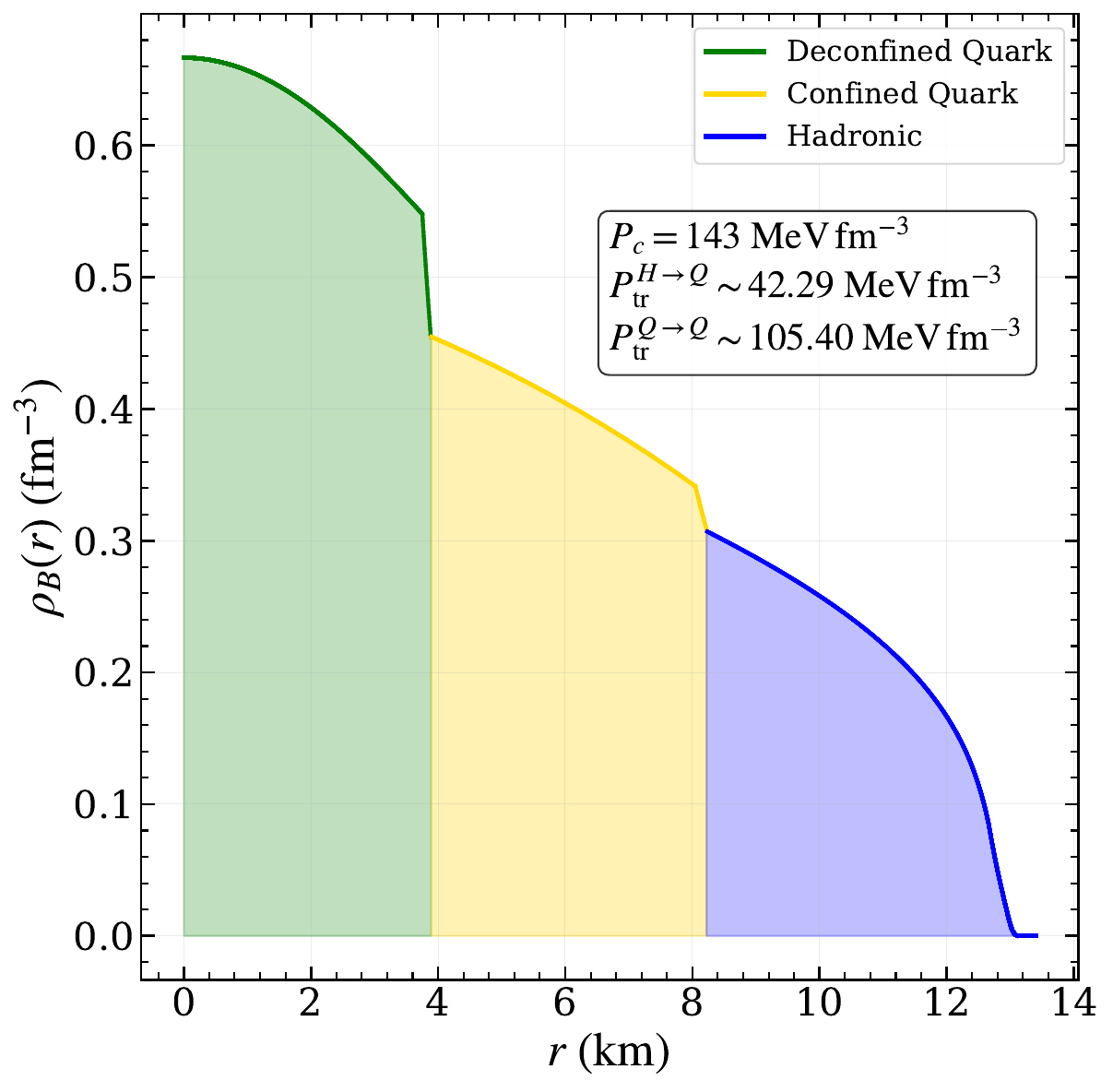}
    \caption{Baryonic density profile as a function of radial distance for the stable maximum mass hybrid star  of Fig.~\ref{fig:HQD_NL3WR}{\color{blue}c}.}
    \label{fig:HQD_density_profile}
\end{figure}

Fig.~\ref{fig:HQD_NL3WR_frac} shows the quark fractions inside the
maximum-mass hybrid star. 
The strange quark fraction becomes nonzero only in the deconfined region. The corresponding baryon density profile of the maximum-mass hybrid star, showing all three phases, is presented in Fig.~\ref{fig:HQD_density_profile}. 
Moving from the center towards the surface, both the deconfined core and the surrounding confined shell extend over a radial distance of about $4$~km each.
\section{Conclusion and outlook}\label{Conc}

We have proposed a modification of the Polyakov-loop potential that gives it an explicit dependence on the quark chemical potential, so that it remains finite in the zero-temperature limit and can therefore be used to describe the confinement-deconfinement transition in cold dense matter, a regime in which the standard PNJL framework loses its order parameter entirely. Combining this modified quark sector with several hadronic models (SFHo, DD2, DD2$_{\rm hyp}$, \mbox{NL3$\omega\rho$}) through a Maxwell construction, we built hybrid-star equations of state and used them to solve the TOV equations for the resulting global stellar properties.

Our systematic scan of the modified Polyakov-loop parameters ($T_0$, $\alpha_0$, $\eta_2$) and of the quark vector couplings ($G_V$, $G_{\v\v}$) shows that each of them shifts the location of the hadron-quark~(\mbox{H-Q}) and quark-quark~(\mbox{Q-Q}, confined-to-deconfined) transitions in a controlled and physically transparent way: increasing $T_0$ or decreasing $\alpha_0$ or $\eta_2$ pushes the Q-Q transition to higher baryonic chemical potential, enlarging the confined window, whereas increasing $G_{\v\v}$ shifts the Q-Q transition to lower density. Increasing $G_V$ or decreasing $B_0$ pushes the H-Q transition to higher density. A result that holds through most of this scan for the low density SFHo EOS is that the  mass-radius sequence resulting from an EOS with the three phases, hadronic, confined and deconfined quark matter, generally becomes mechanically unstable as soon as the H-Q transition is reached. 
However, this is not universal: describing the low density phase with DD2 or \mbox{DD2$_{hyp}$}, the mass-radius becomes unstable at the \mbox{Q-Q} transition. For a further, more finely-tuned corner of parameter space, realized here with the \mbox{NL3$\omega\rho$} hadronic equation of state, we find a genuine, mechanically stable three-phase structure, in which the maximum-mass configuration itself possesses a deconfined quark core,
surrounded by a confined shell, surrounded by a hadronic envelope.

We further find that reaching astrophysically relevant maximum masses with a quark core is not simply a matter of tuning the mPNJL parameters in isolation: with the soft SFHo hadronic equation of state at low density, every quark-core configuration we obtained had a maximum mass below two solar masses, even in the cases where a stable branch existed. The SFHo mass-radius curve has a negative slope over the whole range of masses above one solar mass. In Ref.~\cite{Albino:2025puc}, it was shown that the mass-radius curve of a model that describes two solar mass neutron stars and predicts the appearance of quarks in the center of the star, has with a large probability, a positive slope between $1M_\odot$ and~$\sim 1.4M_\odot$. Replacing SFHo with a stiffer low-density equation of state, keeping sufficiently strong repulsive vector interactions ($G_V$, $G_{\v\v}$), allows all three scenarios to exceed two solar masses: a $\sim 2M_\odot$ hybrid star with a $\sim 4$~km deconfined quark core (DD2), a $\sim 2M_\odot$ hybrid star with a $\sim 5$~km confined quark core, reproduced with two further low-density equations of state (\mbox{DD2$_{hyp}$} and \mbox{NL3$\omega\rho$}), and a $\sim 2.1M_\odot$ hybrid star with a three-phase deconfined--confined--hadronic structure (\mbox{NL3$\omega\rho$}).  For all the three hadronic models, the mass-radius curve shows a backbending with a positive slope for low masses. In the confined quark-core configuration, the speed of sound at the center of the maximum-mass star exceeds the conformal limit, $c_s^2 = 1/3$, consistent with the general expectation that supporting $2M_\odot$ stars with a quark core requires the equation of state to stiffen beyond the naive perturbative-QCD bound. In the deconfined-core and three-phase configurations, by contrast, $c_s^2$ at the center of the maximum-mass star, itself in the deconfined phase in both cases, remains below the conformal limit, showing that the necessary stiffening in these scenarios is achieved in the hadron and confined phases. The deviations from the asymptotic conformal limit in some configurations are physically expected and acceptable at the neutron star densities. 

Taken together, these results show that the \mbox{mPNJL} model introduced here is able to accommodate massive cold hybrid stars with a deconfined core, a confined quark core, or, in a further, narrower corner of parameter space, both simultaneously as a stable three-phase structure, provided a sufficiently stiff low-density hadronic equation of state and sufficiently strong vector repulsion among quarks are used. The present study is deliberately restricted to establishing this qualitative parameter dependence and the physical mechanisms responsible for it, rather than fitting the model to data. A quantitative determination of the preferred parameter ranges through Bayesian inference using current multi-messenger observations, NICER mass-radius measurements, and gravitational-wave tidal deformability
constraints, together with an assessment of the relative statistical evidence for the deconfined-core, confined-core, and three-phase scenarios identified here, and of the conditions under which the three-phase structure is favored, will be presented in a forthcoming study.

\begin{acknowledgements}
This work has been done as a part of the Project INCT-F\'isica Nuclear e Aplica\c{c}\~oes, under No. 408419/2024-5. It is also supported by Conselho Nacional de Desenvolvimento Cient\'ifico e Tecnol\'ogico (CNPq) under Grants No. 307255/2023-9 (O.L.), No. 301779/2025-2 (M.D.), No. 01565/2023-8~(Universal - O.L., M.D.), No. 409736/2025-2~(Universal - O.L, M.D.), No. 444797/2024-6 (O.L., M.D.), and Funda\c{c}\~ao de Amparo \`a Pesquisa do Estado de S\~ao Paulo (FAPESP) under Thematic Project No. 2024/17816-8 (O.L., M.D.), and Thematic Project InTheGra No 2025/12606-8 (O.L., M.D.). This work was partially supported by national funds from FCT (Fundação para a Ciência e a Tecnologia, I.P, Portugal) under project UID/04564/2025, identified by DOI 10.54499/UIDB/04564/2025, and project  2024.16290.PEX identified by DOI  identifier 10.54499/2024.16290.PEX.
\end{acknowledgements}

\bibliographystyle{apsrev4-2}
\bibliography{references}

\end{document}